\documentclass[]{emulateapj}

\newcommand{\lya}{Ly-$\alpha$~}
\newcommand{\lyb}{Ly-$\beta$~}
\newcommand{\lyg}{Ly-$\gamma$~}
\newcommand{\lyd}{Ly-$\delta$~}
\newcommand{\nh}{N_{\mhi}}
\newcommand{\novi}{N_{\movi}}
\newcommand{\nciv}{N_{\mciv}}
\newcommand{\pcmsq}{cm$^{-2}$}
\newcommand{\ovi}{O~{\sc vi}~}
\newcommand{\civ}{C~{\sc iv}~}
\newcommand{\cv}{C~{\sc v}~}

\newcommand{\hi}{H~{\sc i}~}

\newcommand{\heii}{He~{\sc ii}~}

\newcommand{\ov}{O~{\sc v}~}
\newcommand{\ovii}{O~{\sc vii}~}
\newcommand{\oviii}{O~{\sc viii}~}
\newcommand{\oden}{\rho/\bar{\rho}}
\newcommand{\ang}{\AA~}
\newcommand{\mhi}{{\rm H \; \mbox{\tiny I}}}
\newcommand{\mov}{{\rm O \; \mbox{\tiny V}}}
\newcommand{\movi}{{\rm O \; \mbox{\tiny VI}}}
\newcommand{\movii}{{\rm O \; \mbox{\tiny VII}}}

\newcommand{\mciv}{{\rm C \; \mbox{\tiny IV}}}

\begin{document}

\bibliographystyle{/h0/simcoe/latex/apj}

\title{ The Distribution of Metallicity in the IGM at $\mbox{\scriptsize
z}\sim2.5$: \\ \ovi and \civ Absorption in the Spectra of 7
QSOs\altaffilmark{1}}

\author{Robert A. Simcoe\altaffilmark{2,4}, Wallace L.W. Sargent\altaffilmark{2},
Michael Rauch\altaffilmark{3}}

\altaffiltext{1}{The observations were made at the W.M. Keck Observatory
which is operated as a scientific partnership between the California
Institute of Technology and the University of California; it was made
possible by the generous support of the W.M. Keck Foundation.}
\altaffiltext{2}{Palomar Observatory, California Institute of Technology,
Pasadena, CA 91125, USA; ras@astro.caltech.edu, wws@astro.caltech.edu}
\altaffiltext{3}{Carnegie Observatories, 813 Santa Barbara Street,
Pasadena, CA 91101, USA; mr@ociw.edu}
\altaffiltext{4}{Current Address: Pappalardo Fellow, MIT Center for Space Research,
77 Massachusetts Ave. \#37-664B, Cambridge, MA 02139, USA; simcoe@mit.edu}

\begin{abstract}
We present a direct measurement of the metallicity distribution
function for the high redshift intergalactic medium.  We determine the
shape of this function using survival statistics, which incorporate
both detections and non-detections of \ovi and \civ absorption,
associated with \hi lines in high resolution quasar spectra.  The \ovi
sample (taken from 7 QSOs at $z_{abs}\sim 2.5$) contains lines with
$\nh\ge10^{13.6}$, encompassing $\sim 50\%$ of all baryons at
$z\sim2.5$.  Our survey accounts for $\approx 98.8\%$ of the \civ mass
and $\approx 90\%$ of the \ovi mass in the universe at this epoch.  We
find a median intergalactic abundance of $[O,C/H]=-2.82$; the
differential abundance distribution is approximately lognormal with
mean $\left< [C,O/H] \right> \approx-2.85$ and $\sigma=0.75$ dex.  We
discuss the sensitivity of these results to the assumed form of the
ionizing UV radiation field.  Some $\sim 60-70\%$ of lines in the \lya
forest are enriched to observable levels of $[O,C/H]\gtrsim-3.5$,
while the remaining $\sim 30\%$ of the lines have even lower
abundances.  Thus we have not detected a universal metallicity floor
as has been suggested for some Population III enrichment scenaria.  In
fact, we argue that the bulk of the intergalactic metals formed later
than the first stars that are thought to have triggered reionization.
We do not observe a strong trend of decreasing metallicity toward the
lower density IGM, at least within regions that would be characterized
as filaments in numerical simulations.  However, an [O/H] enhancement
may be present at somewhat high densities.  We estimate that roughly
half of all baryons at these redshifts have been enriched to
$[O,C/H]\gtrsim-3.5$.  Using a simple ``closed box'' model for the
metallicity evolution of the IGM, we estimate the chemical yield of
galaxies formed prior to $z\sim 2.5$, finding that the typical galaxy
recycled $0.1-0.4\%$ of its mass back into the IGM as heavy elements
in the first 3 Gyr after the Big Bang.

\end{abstract}


\section{Introduction}\label{sec_introduction}

Observations of \civ absorption in QSO spectra have unambiguously
revealed the presence of heavy elements in the \lya forest at $z=3$
\citep{meyer_york,cowie_civ_1,ellison_civ, cowie_civ_2,rauch_civ,
rauch_civ_2, schaye_civ_pixels}.  These results have been interpreted
as evidence of widespread enrichment of the universe with the
byproducts of stellar nucleosynthesis, but the data at present permit
the possibility of several enrichment mechanisms, from an early
($z>10$) generation of massive short-lived stars
\citep[e.g.,][]{ostriker_gnedin} to more recent ($z<7$) pollution by
winds from star forming galaxies
\citep{kurt_winds,aguirre_outflows,springel_outflows}.  Measurements
of the IGM metallicity can usefully constrain the history of star
formation and stellar recycling on cosmological scales.
However, at a limiting associated \hi column of $\sim 10^{14.5}$
\pcmsq, existing \civ measurements only probe regions with gas
overdensities of $\rho/\bar{\rho}\ge 7$ relative to the cosmic mean.
This corresponds to the densest $\sim 15\%$ of the baryons in the \lya
forest, which occupy $\lesssim 1\%$ of the total volume of the
universe.

At \hi column densities below $10^{14.5}$ \pcmsq, \civ searches are
limited by signal-to-noise ratio even in the best available spectra.
A decline in the overall carbon column in low density gas is
exacerbated by the increasing ionization state of the gas, which
favors \cv and higher levels in the more tenuous regions of the IGM.
To combat this effect, several studies have employed sensitive
statistical techniques to extract averaged estimates of [C/H] at lower
densities.  These studies generally fall into two categories: either a
shift-and-stack approach to generate a high signal-to-noise, composite
\civ line \citep{limin_civ}, or a pixel-by-pixel statisical analysis
of the relative optical depths of \hi and \civ throughout the run of a
spectrum \citep{cowie_civ_nature, songaila_civ, ellison_civ,
aguirre_civ}.  The reader is referred to \citet{ellison_civ} for a
review of these two methods.  Over time, pixel methods have seen more
extensive use, although they can be difficult to interpret without
recourse to simulations, and in our experience they have been somewhat
sensitive to slight systematics (e.g. continuum placement).
\citet{schaye_civ_pixels} and \citet{aguirre_siiv_pixels} have
recently compiled comprehensive studies of pixel statistics which
address many of these issues, using numerical simulations to trace the
distribution of intergalactic carbon and silicon with density and
redshift.

In this work, we describe a new effort to pursue metallicity
measurements in the IGM to near the cosmic mean density through
observations of weak \ovi absorption in a sample of 7 QSOs.  This is
supported by new observations of \civ absorption in high signal to
noise spectra of two sample objects.  We have limited our measurements
to a single observed quantity: the column density of detected \ovi and
\civ lines (or $3\sigma$ upper limit on $\novi,\nciv$ for
non-detections).  This paper therefore represents our best effort to
describe the enrichment of the IGM using the direct, local properties
of individual systems.  This approach accepts a slight sacrifice in
sensitivity over the statistical methods quoted above, in return for a
localized description of the metallicity field and a straightforward
assessment of possible systematic biases.  It also allows for an easy
division of the total sample into subsets to study trends of
metallicity with other external variants, such as \hi column density,
or proximity to galaxies (though the latter is beyond the scope of
this paper).  

The literature contains several examples of intergalactic \ovi searches,
as it has long been known that its ionization potential and 
high abundance are very favorable for production in the low density 
IGM \citep{norris1983,chaffee86}  
\footnote{N.B.: Searches for weak metals in the high redshift \lya
  forest are qualitatively different from the recent low redshift \ovi
  surveys undertaken with FUSE and STIS
  \citep{tripp_ovi_galaxies,tripp_ovi_metals,
  savage_fuse_ovi_summary}.  The local, ``warm-hot'' variety of \ovi
  is usually collisionally ionized through interactions with galaxies,
  or by accretion shocks as pre-enriched material falls onto large
  scale structure.  The weaker high redshift population should be
  predominantly cooler gas that is photoionized by the integrated
  light from quasars \citep{rauch_simulations,carswell2002}, although
  we reported in \citet{simcoe2002} on a separate, co-existing
  population high redshift systems that may be collisionally ionized.
  Photoionized \ovi systems are physically distinct from the
  ``warm-hot'' phase of the IGM \citep{dave_whim}. }.
The principal hurdle faced by these searches is the unfortunate rest
wavelength of the \ovi doublet ($1032,1037\AA$).  This locates the
weak \ovi lines deep in the \lya forest, which smothers the signal of
many real systems and further introduces a number of false positive
identifications.  Many of the \ovi systems discovered in these surveys
display large equivalent widths and rich chemical structure, as would
be expected in galactic environments
\citep{burles_ovi_survey,simcoe2002}.  Using a single spectrum of
Q1422+2309, \citet{dave_ovi} statistically explored the \ovi content
of the low density IGM at $z\sim3.5$, finding evidence for declining
abundances in lower density regions.  However,
\citet{schaye_ovi_pixels} have claimed a detection of \ovi even at
very low densities, reaching below the cosmic mean.  More recently
\citet{carswell2002} and \citet{bergeron_ovi} have discovered evidence
for \ovi enrichment in a number of individual systems with
$\nh\sim10^{14.5}$ at $z\sim 2.3$.

\begin{deluxetable}{c c c}
\tablewidth{0pc}
\tablecaption{Object Summary}
\tablehead{{Object} & {$z_{em}$} & {$\Delta z_{\small OVI}$\tablenotemark{1}}}

\startdata
Q1009+2956 & 2.62 & 2.295-2.553       \\
Q1217+4957 & 2.70 & 2.374-2.635       \\ 
Q1347-2457 & 2.53 & 2.329-2.525       \\
Q1442+2931 & 2.63 & 2.270-2.556$^2$   \\
Q1549+1919 & 2.83 & 2.273-2.767       \\
Q1603+3820 & 2.51 & 2.201-2.415$^3$   \\
Q1700+6416 & 2.72 & 2.259-2.651       \\
\enddata

\tablenotetext{1}{\small Corrected to exclude regions within 5000 km/s of the
QSO emission redshift.}
\tablenotetext{2}{\small The region $2.41<z<2.46$ was excluded for
  this object due to the presence of a weak Damped \lya system in 
  this wavelength range.}
\tablenotetext{3}{\small The region $2.415<z<2.450$ was excluded for
  this object due to the presence of a strong system which appears to
  be ejected from the QSO.}
\end{deluxetable}

\label{tab_observations}

This paper expands upon these studies using an increased number of
sightlines observed at high resolution, and also by incorporating the
full use of non-detections in the analysis.  These additional
measurements have allowed us to use the methods of survival analysis
to construct the distribution function of [O/H] and [C/H] in the \lya
forest.  We shall demonstrate that our measured distribution function
agrees quite well with very recent models produced from the completely
different method of pixel statistics \citep{schaye_civ_pixels}.  In
Section \ref{sec_observations} we describe the observations, sample
definition, and measurement methods; Section \ref{sec_analysis}
explains the conversion of column density measurements to abundance
estimates and the application of survival statistics to estimate the
[O/H] distribution; Section \ref{sec_discussion} discusses the
cosmological implications of the observed metallicity distribution.

\section{Observations}\label{sec_observations}

Our search targets 7 bright QSOs in the redshift range $2.5<z<2.8$,
which was chosen to balance three important factors regarding the
existence and observability of \ovi: contamination from the \lya
forest (which worsens toward higher redshift), strength of the
metagalactic UV ionizing flux (which is maximized at $2.5<z<3.0$), and
accessibility from large aperture, ground-based telescopes (to improve
S/N for the weakest lines).  The sightlines are listed in Table 1,
along with their QSO emission redshifts and the range of redshift
covered by the absorption line measurements.  Four of the seven
spectra were used in our earlier analysis of strong \ovi systems
\citep{simcoe2002}, and the details of those observations are found
therein.  New observations of Q1217+4957, Q1347-2457, and Q1603+3820
were taken in April 2002 under variable conditions; Q1626+6433 was
used in the previous paper but is omitted here because of the data's
lower signal to noise ratio and small redshift coverage.  The
observations were taken with the HIRES spectrograph on the Keck I
telescope, using the UV blazed cross disperser.  All exposures were
taken through an $0\farcs86$ slit fixed at the parallactic angle, for
a spectral resolution of 6.6 km s$^{-1}$, and the data were reduced
using T. Barlow's ``makee'' echelle reduction package.

\subsection{Identification and Measurement of the \ovi Systems}

Because \ovi is placed within the \lya,\lyb, and higher order Lyman
forests, our first step has been to follow the procedure of
\citet{carswell2002}, fitting the entire \lya forest region to remove
high order \hi transitions from the data.  Beginning at the emission
redshift of the quasar, each \hi line in the forest was fit with a
combination of Voigt profiles using the VPFIT \footnote{Provided by
R. Carswell, J. Webb, A. Cooke, \& M. Irwin -
http://www.ast.cam.ac.uk/$\sim$rfc/vpfit.html} software package.  This
procedure was extended to lower redshift until the observed wavelength
of \lya was equal to the observed wavelength of \ovi at the emission
redshift of the QSO.  For the handful of cases in each spectrum with
$\nh>10^{15}$ (where \hi lies on the flat part of the curve of growth)
we performed joint fits for the \hi column density using
\lya,\lyb,\lyg, and \lyd.  In all other cases, only \lya was used, as
the primary goal was to remove these higher order lines.  As a side
benefit, we obtain a full statistical description of the forest in the
redshift range of the \ovi survey.  This information is used in
Section 3.7 to estimate the mass fraction of baryons probed in the
sample.

Once a satisfactory fit was obtained for the forest, we
adjusted the original data and error arrays, using the model fit to
remove the signal of \lyb and higher order Lyman series transitions.  
Even small fluctuations around the continuum level were often caused 
by \lyg to Ly-7 transitions from higher redshift systems.  
By removing this signal the useful \ovi pathlength in each
spectrum is nearly doubled, correspondingly doubling the size of 
our line sample.  

\subsubsection{\ovi Sample Selection and Measurements}

For the \ovi measurements, we have focused on the subset of systems
with \hi column densities above $\nh=10^{13.6}$ \pcmsq.  This
represents almost an order of magnitude increase in \hi depth over
past \civ and \ovi surveys, and corresponds to clouds with
overdensities of $\rho/\bar{\rho}\sim 1.6$ relative to the cosmic mean
at $z\sim 2.5$.  The quality of the \ovi data varies throughout the
survey according to sightline and redshift, and not all regions are
sufficiently sensitive to reveal \ovi lines at the level expected for
the weakest \hi systems in the sample.  However, in some regions the
data quality is very high, and we shall describe below how
measurements from the best regions can be combined consistently with
upper limits measured in portions of the spectrum with lower
signal-to-noise ratios.

We selected candidate systems using \hi column densities obtained from our
Voigt profile fits to the \lya forest.  For each line in the
$\nh\ge10^{13.6}$ sample, we examined the spectrum (now free
of higher order \hi lines) for \ovi at each sample redshift.  
In instances where a possible \ovi line was detected at $\ge 3\sigma$
significance, we used VPFIT to determine the column density and $b$ 
parameter of the \ovi line.  At this stage, systems were flagged as 
possible detections if both components of the \ovi doublet were
clearly visible, or if
absorption was present in one component but the other was strongly
blended with a \lya forest line.  If no \ovi absorption was present at the
expected location of one or both doublet components, we determined
$3\sigma$ upper limits on the \ovi absorbing column.  The upper
limits were measured from the \ovi 1032\AA ~transition except in cases
where \lya forest blending led to a cleaner result from \ovi1037\AA.

When no \ovi is detected, the measurement of an upper limit on $\novi$
depends upon the choice of linewidth, $b$, which cannot be determined
from the data.  Rather than guessing at the correct value to use, we
have measured two upper limits for each system.  For one measurement,
we fix $b_{\movi}$ to the value for completely thermal line broadening
in the \hi-\ovi gas mixture:
$b_{\movi}=b_{\mhi}\times\sqrt{m_\mhi/m_\movi}$.  The other limit
correponds to broadening from turbulent or bulk gas flows:
$b_{\movi}=b_{\mhi}$.  The actual value should lie between these two
extremes.  We also fix $z_\movi=z_\mhi$ for measuring limits on
non-detections; for detections we allow the redshift to vary up to
$\sim 40$ km s$^{-1}$ to optimize the fit.  Often velocity differences
are observed between $z_\mhi, z_\movi,$ and $z_\mciv$ in strong
systems.  However, for our non-detections the upper limits are not
strongly sensitive to the exact placement of $z_\movi$, as the noise
properties of the data tend to be fairly uniform over the $\sim 20$ km
s$^{-1}$ velocity offsets sometimes observed between different ions.
All quoted upper limits are at the $3\sigma$ level when no line could
be fit.  In cases where VPFIT was able to fit a line with $N_{vp}$ at
$\le 3\sigma$ significance, we quote an upper limit of $\novi\le
N_{vp} + 3\sigma$.  Regions where VPFIT found absorption at the
$3\sigma$ level or greater are reported as detections with the best
fit column density.

\begin{deluxetable}{c c c}
\tablewidth{0pc}
\tablecaption{Numbers of \ovi Detections vs. Non-Detections}
\tablehead{{Class} & {$N_{\rm thermal}$} & {$N_{\rm turb}$}} 
\startdata
$dd$ & 41  & 41 \\
$bd$ & 71  & 79 \\
$nd$ & 118 & 110 \\
\enddata
\end{deluxetable}\label{tab_detections}

In most of our systems, one component of the \ovi doublet is blended with 
a \lya forest line.  If the other transition is clean of blends and shows 
no \ovi its interpretation as a non-detection is secure.  This
situtation is fairly common, and we denote it as a $bn$ system, for 
``blend/non-detection''.  But often one transition is blended with a
forest line while the other shows weak absorption that could be 
either \ovi or another interloping \lya line.  These systems,
hereafter called the $bd$ sample (``blend-detection''), comprise 
a potentially serious source of false positive \ovi identifications.  
A conservative approach is to treat all $bd$s as upper limits due to 
their unsure identification.  However, if we assign upper limits to 
all $bn$s and $bd$s we would downweight many true detections 
and bias our results toward low metallicities.  Yet if we were to 
restrict our sample to those systems where both components of the
doublet are completely free of blending (accordingly, the $dd$ 
detections and $nn$ nondetections) the sample would be too small to 
provide physically interesting constraints on the metallicity of the
IGM.  At the measurement stage, our approach has been to treat 
all $bn$ and $nn$ systems as upper limits, and all
$bd$ and $dd$ systems as detections for completeness.  Later in the 
analysis, we use Monte Carlo simulations to estimate the degree
of contamination from false positives in the $bd$ subsample, and
correct for their effects statistically.

\subsection{Identification and Measurement of the \civ Systems}

For two of the objects in the sample, we have obtained particularly
high quality spectra to search for extremely weak \civ absorption 
features.  Our intent is to compare the distribution of [C/H] with
that of [O/H], both to check for consistency and to test the accuracy
of the ionization corrections that we will use to estimate metallicities.  
By far the best spectrum in the sample is that of Q1549+1919, with an 
average signal to noise ratio of $\sim 350$ per resolution element 
throughout the \civ region.  We also analyze the \civ absorption in 
Q1700+6416, whose spectrum has $S/N\sim 175$ per resolution element.  
While the second spectrum is not as sensitive for
measuring the weakest \civ lines, we have included it to increase the \civ
pathlength and improve the statistics of stronger absorbers.  

As before, we use Voigt profile fits to the \lya forest to
determine the subset of lines to be examined for \civ absorption.
However, even with such high quality data the predicted \civ strength drops 
so precipitously with $\nh$ that we cannot probe overdensities as low as
with \ovi.  Therefore, we limit the \civ sample to systems with \hi 
column densities above $\nh=10^{14.0}$ \pcmsq.  Because the same 
objects were observed for \ovi, we were able to fit across the
Lyman series for saturated \hi lines
so the \hi columns are very accurately known.  
At each sample redshift, we examine the location of the 
\civ $1548.202$\ang line and measure its strength if a detection is made.
If no feature is present, we determine a $3\sigma$ upper limit on the \civ
column density for both the thermal and turbulent broadening cases.  
For nearly all of the systems, we used the $1548.202$\ang component to measure
upper limits.  In the unusual case where this line was blended 
with a line from another redshift system, we used the $1550.774$\ang 
component to measure the limits.

\subsection{Distribution in the \hi/\ovi and \hi/\civ Planes}

\begin{figure}
\plotone{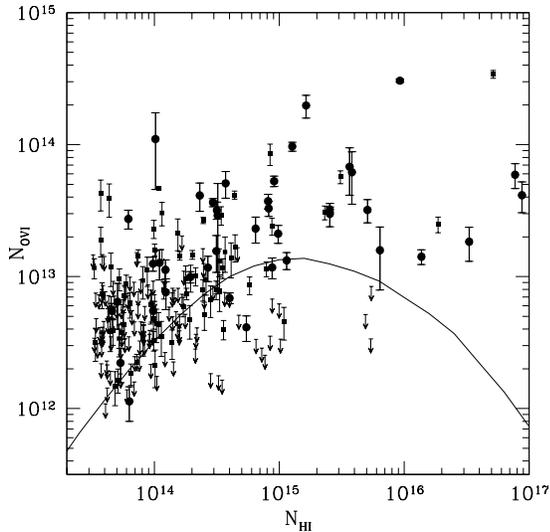}
\caption{ 
  Scatter plot of detections and upper limits in the $\nh/\novi$
  plane.  Solid round points with errors represent the $dd$ variety of
  \ovi detections, and small squares represent $bd$ detections.  Arrows 
  indicate $3\sigma$ upper limits for undetected, thermally broadened
  \ovi.  The solid line 
  drawn through the data represents the expected locus of points for 
  a mean $[O/H]=-2.5$ and our fiducial UV background spectrum.
}
\label{fig:ovihi}
\end{figure}

\begin{figure}
\plotone{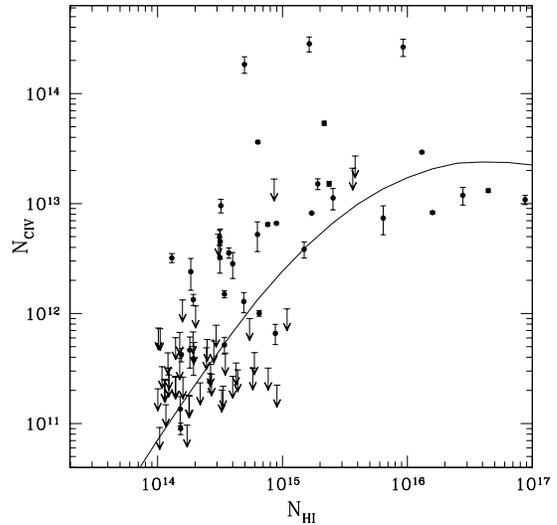}
\caption{ Scatter plot of \civ and \hi data.  Data from only two
  sightlines (Q1549+1919 and Q1700+6416) are used.  Model curve is as
  in Figure \ref{fig:ovihi}.  }
\label{fig:civhi}
\end{figure}

We have measured \ovi and \hi column densities for a total of 230 \lya
forest lines with $\nh\ge 10^{13.6}$ \pcmsq ~in 7 sightlines, and \civ
and \hi column densities for a total of 83 systems with $\nh\ge
10^{14.0}$ \pcmsq ~in two sightlines.  These data are summarized in
Figures \ref{fig:ovihi} and \ref{fig:civhi}, where we plot
measurements or $3\sigma$ upper limits for each \ovi and \civ line
against its \hi column density.  The upper limits shown in the plots
are for the thermal line broadening case.

In the \ovi figure, we have subdivided the detected systems according
to our confidence in their identification.  Heavy round points denote
highly probable \ovi, where both components of the doublet are seen in
the proper ratio of strengths (the $dd$ sample).  Smaller squares are
used for the $bd$ sample, where only one of the two doublet components
could be measured due to \lya forest blending with the other line.
Again, we stress that this subsample contains spurious measurements
from interloping \hi lines - in Section 3.5 we estimate that only
$\sim 40\%$ of these points are actually \ovi.  In false positice
detections the measurements still provide upper limits on the \ovi
column density, so in truth some $\sim60\%$ of the square points would
be replaced with upper limit symbols.  The \civ sample does not suffer
from this sort of contamination, and all detections are weighted
equally.

The solid curve in each panel indicates the locus
of points predicted for \lya forest clouds enriched to 
[O/H]=[C/H]=-2.5
\footnote{We use the standard notation where $[X/H]$ represents the
log of the abundance for element X relative to the solar level, i.e.
$[X/H]=\log(X/H)-\log(X/H)_\sun$.}  , and illuminated by a slightly
modified version of the \citet{HM96} ionizing background spectrum (see
Sections 3.1 and 3.2 for a description of how these model predictions
are generated).  Our observations have reached the sensitivity
required to detect metal lines at this commonly quoted abundance level
for much of the sample.  Even at $\nh<10^{14.0}$, some of our \ovi
measurements are able to meaningfully probe the metal distribution,
although the presence of several upper limits above the model curve
indicates the increasing challenge at the lowest $\nh$.

Generally, the \ovi and \civ data points follow the trends traced by
the model predictions, with a few exceptions.  At high column
densities (about $\nh>2\times 10^{15}$) most \hi systems appear to
contain \ovi, but the \ovi is often significantly stronger than
predicted.  This discrepancy is not surprising, and could be due to
several factors.  First, the recipe for producing the model curve
assumes a scaling between physical gas density and \hi column which
may break down at higher densities where cosmic structure becomes
nonlinear.  More importantly, the models calculate metal strengths
assuming a pure photoionization equilibrium, whereas the strongest
systems may be collisionally dominated.  Finally, the dense systems
are likely sites of local chemical enrichment so their metallicity
could be higher by an order of magnitude or more.

At lower $\nh$ there are several systems with exceptionally {\em low}
metal to \hi ratios, both for \civ and \ovi.  In some cases these lie
almost an order of magnitude below the model prediction, even though
the model should be most accurate in this regime.  The presence of
several such systems was initially a puzzle as we expected to find
\ovi and \civ absorption throughout the IGM.  It is not obvious from
Figures \ref{fig:ovihi} and \ref{fig:civhi} whether these outlying
systems are consistent with simple scatter about a trend, or whether
they represent a different population of lines that is statistically
distinct from the rest of the sample.  This question will be addressed
in detail below as we construct a quantitative distribution function
of the intergalactic metallicity field.

\section{Analysis}\label{sec_analysis}

Having fit \ovi and \civ line strengths for the \hi selected
systems, we now translate these measurements into metallicity estimates
for each line in the sample.  In the following sections we describe
the methods used to build the metallicity distribution using survival 
statistics, including corrections for false positive \ovi identifications.

\subsection{Estimating [O/H], [C/H] for \lya Forest Lines}\label{sec:metallicity}

Using measurements of \ovi and \hi, the oxygen abundance for a
single system is calculated as:
\begin{equation}
\left[{{O}\over{H}}\right] = \log\left({{{\novi}\over{\nh}}}\right) +
\log\left({{f_{\mhi}}\over{f_\movi}}\right) - 
\log\left({{O}\over{H}}\right)_\sun,
\label{eqn:metallicity}
\end{equation}
\noindent where $f_\movi=n_\movi/n_{\rm O}$ and $f_\mhi=n_\mhi/n_{\rm
H}$ are ionization fractions and we have assumed solar relative
elemental abundances \footnote{In Section \ref{sec:final_Z} we discuss the
implications of other choices for [C/O].}.  Clearly the same method may
be applied to calculate [C/H] from our \civ line sample.  Also, the
equation may be inverted to predict $\novi$ for different values of
$\nh$ if [O/H] is already known (this is the method used to produce
the model curves in Figures \ref{fig:ovihi} and \ref{fig:civhi}).

Throughout, we use the meteoritic solar abundances of
\citet{grevesse_solar_abund}, with $A_{\rm oxygen}=8.83$ and $A_{\rm
carbon}=8.52$ on a scale where $A_{\rm hydrogen}=12$.  Several
revisions to these Solar abundances have been suggested since the
publication of \citet{grevesse_solar_abund}, including
\citet{allende_prieto_oc}, \citet{allende_prieto_oxygen},
\citet{asplund_solar_abundances}, and
\citet{holweger_solar_abundances}.  Allende Prieto advocates a
downward revision of both the carbon and the oxygen solar abundances
by $0.13-0.14$ dex.  Likewise, Asplund reduces $A_{\rm oxygen}$ by
0.17 dex and $A_{\rm carbon}$ by 0.11 dex.  Holweger uses a standard
oxygen abundance that is $0.09$ dex lower, but a carbon abundance that
is $0.07$ dex higher.  We have continued to use the
\citet{grevesse_solar_abund} abundances for ease of comparison with
prior studies of intergalactic metal enrichment.  Readers who prefer
the Allende Prieto or Asplund values should increase all [C/H] and
[O/H] estimates in this paper by $\approx 0.13-0.14$ dex.
Renormalization of our measurements to Holweger's scale is slightly
less straightforward.  It requires a similar additive adjustment to
the solar zero point, but also a hardening of the UV background field
to match the relative distributions of carbon and oxygen (see section
\ref{sec:final_Z}).  The net effect is a downward shift in all of the
metallicities listed throughout the paper, by $\approx 0.07-0.1$ dex.

\subsection{Modeling Intergalactic Ionization Conditions}\label{sec:ionization_model}
\begin{figure}
\plotone{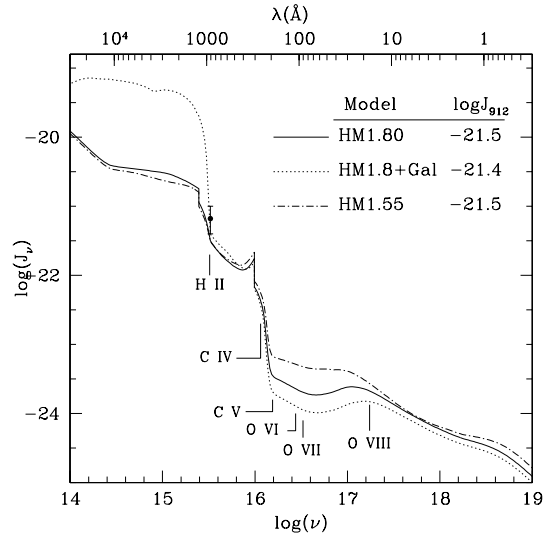}
\caption{Three representative models of the UV/X-Ray background
  used in the analysis.  The solid line represents our favored HM1.80 model.
  The ionization edges for several important
  transitions are labeled, with the convention that the ``\ovi'' label
  marks the wavelength associated with the ionization of \ov to \ovi.  
  The point with errorbars represents the \citet{scott_uv_bkgd}
  estimate of $\log J_{912}$ from the proximity effect.
}
\label{fig:uvbg}
\end{figure}

The principal challenge of our metallicity calculation is an accurate
estimation the ionization correction term in Equation
\ref{eqn:metallicity}.  We have run simulations of the ionization
conditions in the IGM to calculate this term as a function of \hi
column density for each line in the sample.

In the low density \lya forest, simulations suggest that there exists
a tight correlation between physical gas density and observed \hi
column density \citep{zhang_forest, dave_forest, schaye_forest}.  We
can use this relation to translate each \hi column density in the
sample into a physical density, inverting Equation 8 of
\citet{schaye_forest}:
\begin{eqnarray}
n_{\rm H} \approx 10^{-5} {\rm cm}^{-3}\left({{\nh}\over{2.3\times 10^{13}
    {\rm cm}^{-2}}}\right)^{\frac{2}{3}} ~~~~~~~~\nonumber \\ 
\times T_4^{0.17}
    \Gamma_{12}^{\frac{2}{3}} \left({{f_g}\over{0.16}}\right)^{-\frac{1}{3}}.
\label{eqn:schaye}
\end{eqnarray}
The last three factors represent scaling corrections for the gas
temperature, ionizing background, and fractional mass in diffuse gas,
and are likely to be of order unity as argued by Schaye.  This
relation was derived assuming that the local collapse of \lya forest
clouds occurs nearly in hydrostatic equilibrium.  It may break down at
high densities where the gas becomes optically thick ($\nh\gtrsim
10^{17}$), or at $\oden < 1$ where gravitational timescales become
comparable to the Hubble time.  It has not been verified
observationally over most of the $\nh$ range considered here, since
\lya forest systems typically exhibit few or no other lines which
could be used to construct an estimate of $n_{\rm H}$.  However,
numerical investigations indicate that the scaling remains accurate
from roughly the mean density to $\nh\lesssim 10^{16}$
\citep{dave_forest}.

The intergalactic gas is highly ionized by ambient ultraviolet/X-Ray
radiation, whose spectrum at $z\sim2.5$ is dominated by quasar light
that is reprocessed by radiative transfer through the IGM.  We have
run sets of ionization calculations for several different forms of the
ionizing background spectrum, which were computed as described in
\citet{HM96} and kindly provided to us by F. Haardt.  A range of these
spectra is shown in Figure \ref{fig:uvbg}; details about their
construction are provided in Appendix A.  The curves differ both in
their normalization, and in their underlying UV spectral slope.  The
discrepancies reflect uncertainty in the exact shape of the source
function in the radiative transfer equations - i.e., the average UV
spectral slope intrinsic to QSOs.  All of the models assume that
quasar spectra in the UV are pure power laws:
$F_\nu\propto\nu^{-\alpha}$.  Intergalactic \hi and \heii absorption
and reemission modulate this raw spectrum, producing the features seen
in the $10-1000$\ang range of Figure \ref{fig:uvbg}.  We have named
the models according to their input spectral slope, so that the HM1.80
model represents a spectrum with an input $\alpha=1.8$ power law that
has been propagated through the IGM.  The overall normalizations of
the spectra are specified at 912\ang (1 Ry) as indicated in the plot.

\begin{figure}
\plotone{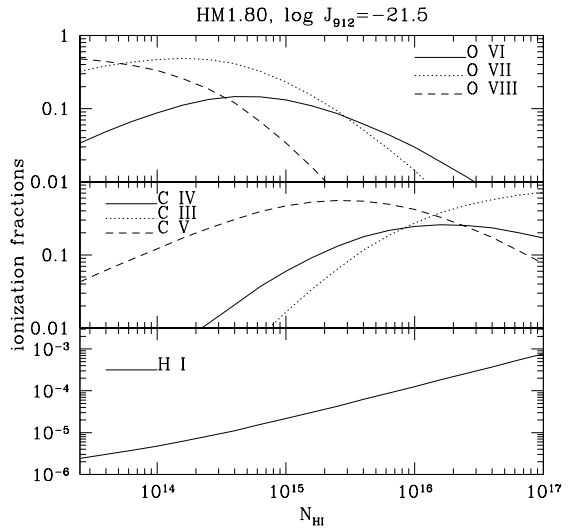}
\caption{The ionization balances of several carbon and oxygen
  transitions relevant to \civ and \ovi production, as a function of
  \hi column density.
}
\label{fig:ion_fracs}
\end{figure}

Recent attempts to measure $\alpha$ by averaging quasar spectra have
produced varied estimates in the range $1.5<\alpha<1.8$.  The original
\citet{HM96} paper used an $\alpha=1.5$ model.  Subsequently,
\citet{zheng_uvbg} published a revised estimate of $\alpha=1.8$, based
on FOS observations of 101 quasars.  The most recent determination of
the UV slope \citep[184 QSOs]{telfer_uv_bkgd} revises the value again
to $\alpha=1.57$, but raises the possibility of strong variations from
object to object.  Each of these papers used the entire sample of FOS
quasar spectra in the HST archive at the time of its publication.  The
samples are therefore not uniformly chosen, and could be subject to
selection effects.  We have tested each of the different QSO models,
and also a HM1.80 QSO spectrum with additional \hi ionizing flux from
galaxies.  For consistency below we adopt the HM1.80 quasar spectrum
normalized to $J_{912}=-21.5$ (solid line in Figure \ref{fig:uvbg}) as
our fiducial X-Ray/UV background.  We shall show that this choice
results in the best match between the intergalactic distributions of
[O/H] and [C/H].  It is also roughly consistent with observations of
the proximity effect and X-Ray background at similar redshifts
\citep{scott_uv_bkgd,boyle_xrlf}.

Using these input spectra, we simulate the ionization balance of the
IGM over a range of densities using the CLOUDY software package
\citep{cloudy}.  We treat the absorbing material as a plane-parallel
gas slab, and CLOUDY calculates the ionization balance for each
element, from which we extract the ionization fractions $f_\movi$,
$f_\mciv$, and $f_\mhi$ (Shown in Figure \ref{fig:ion_fracs}).  These
are combined to calculate the second term in Equation
\ref{eqn:metallicity} along a grid of \hi points.  The exact
ionization correction for each absorption line in the sample is then
obtained by interpolating its \hi column within the grid, and we
combine these with the $\novi, \nciv$, and $\nh$ measurements from the
data to calculate a value or $3\sigma$ upper limit for [O/H] or [C/H].

\subsection{Metallicites of individual lines, and trends with $\nh$}\label{sec:metal_scatter}

\begin{figure}
\plotone{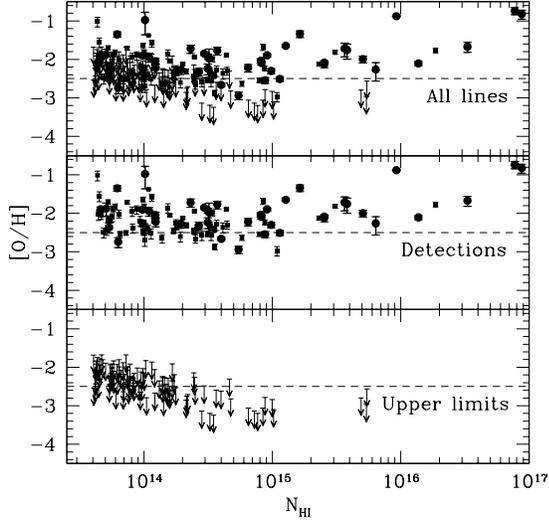}
\caption{Scatter plot of individual metlalicity measurements for all
  \ovi lines in the $\nh>10^{13.6}$ sample.  In the bottom two panels
  the detections and non-detections are shown seperately for clarity.
  The dotted line drawn at [O/H]=-2.5 is not a fit to the data. 
  The \ovi detections exhibit a fairly constant metallicity with
  decreasing density.  The slanted trend in the upper limits is purely
  due to the detection threshold of the data.  In the most sensitive 
  region between $10^{14}<\nh<10^{15}$, there is more scatter in the
  upper limits toward low metallicity, revealing the presence of a
  population of low metallicity systems.  
}
\label{fig:ovi_metal_scatter}
\end{figure}

\begin{figure}
\plotone{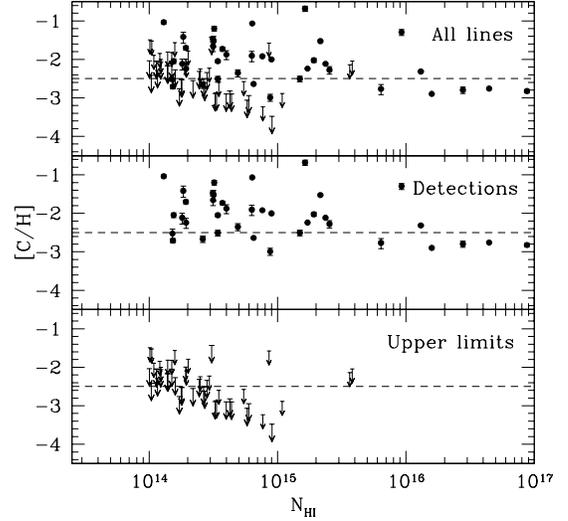}
\caption{
  Scatter plot of individual metallicity measurements for all
  \civ lines in the $\nh>10^{14.0}$ sample.  Labels are as in Figure \ref{fig:ovi_metal_scatter}.
}
\label{fig:civ_metal_scatter}
\end{figure}

In Figures \ref{fig:ovi_metal_scatter} and
\ref{fig:civ_metal_scatter}, we show the metallicity or its upper
limit for each line in the \ovi and \civ samples, plotted against \hi
column density.  In this plot and those that follow, we have assumed
the HM1.80, $\log J_{912}=-21.5$ fiducial spectrum for the UV
background.  The top panel includes all lines in each sample, and in
the bottom two panels we have plotted detections and non-detections
separately for viewing clarity.  The dashed horizontal lines are drawn
at an abundance of $[O/H]=-2.5$ to guide the eye, and are not fits to
the data.

Looking first at the detections, we see that over a broad range in \hi
density there is little change in the average metal abundance.  At the
highest densities there is a clear trend toward increasing metallicity
in the \ovi sample, with no corresponding change in the \civ
distribution.  As discussed above, this could be caused by local
enrichment, but is also likely due in part to model errors.  In
particular, at high densities ($\nh \gtrsim 10^{15.5}$) many of the
systems will be multiphase absorbers with hot collisionally ionized
\ovi and cooler, photoionized \civ \citep{simcoe2002}.  Thus at these
densities the \ovi and \civ might not be expected to follow each other
too closely, and our photoionization models may break down.
Nonetheless, at lower densities where the models are most accurate and
where most of the lines are found, the metallicities of the detected
systems are fairly uniform.  To the limits of the survey
($\rho/\bar{\rho}\sim 1.6$) {\em we have detected no strong trend of
decreasing metallicity toward the more tenuous regions of the IGM}.

This statement is based only upon the detected systems; the upper
limits in the bottom panel reveal a different central result: that
{\em a number of systems have metallicities significantly below the
trend outlined by the detected systems}.  This is most easily seen in
the $10^{14.5}<\nh<10^{15.0}$ range where our measurements can best
sample the scatter both above and below the mean.  In this regime most
of the detected points fall near $[O/H]\sim-2.5$, but there are many
non-detections whose $3\sigma$ upper limits are an order of magnitude
lower, with little or no corresponding scatter at high metallicity
(this is perhaps best seen in the top panel of Figure
\ref{fig:ovi_metal_scatter}).  This pattern is likely to continue at
lower densities, though we cannot test this hypothesis with the
present data due to its limited signal-to-noise ratio (seen as a
decrease in the sensitivity of the upper limits at low $\nh$).

It could be that some of our lowest limits are caused by an occasional
failure of the density and/or ionization model developed above.  A low
metallicity results when we assume that much of the oxygen is in the
\ovi state.  If the gas is more highly ionized (e.g. through shock
heating) or less ionized (e.g. if Equation \ref{eqn:schaye}
underestimates $n_H$) our [O/H] estimate may be erroneously low.  This
effect should not arise from purely intrinsic scatter about the
relation in Equation \ref{eqn:schaye}; in simulations the scatter's
amplitude is only $\sim\pm 0.05$ dex in $n_H$, which translates to
$\sim \pm 0.03$ dex in [O/H].  However, we cannot discount the
possibility of the occasional pathological case that breaks the model.
By studying a large sample of lines we hope to minimize the effects of
individual strange systems.

\begin{figure}
\plotone{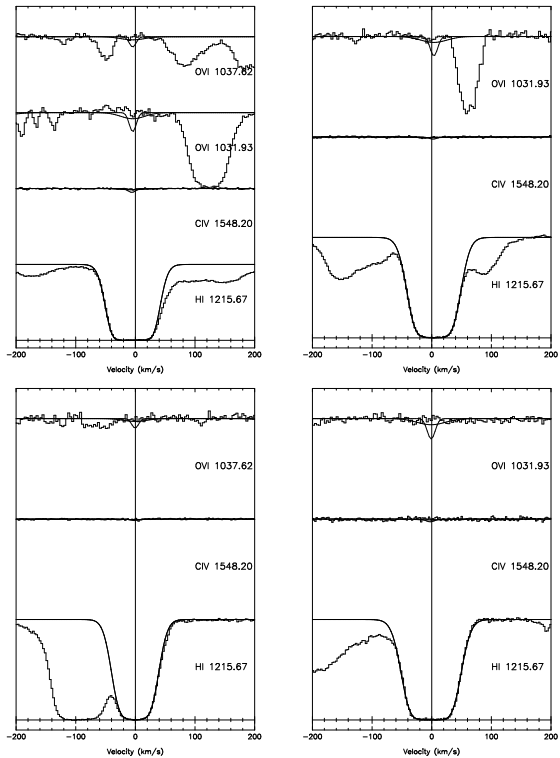}
\caption{ 
A montage of low metallicity ``$nd$'' systems.  The \hi \lya transition is
shown in the bottom of each panel, along with the \civ 1548\AA
~transition and one or both \ovi transitions.  The continuous curves
show the predicted strength of \ovi and \civ if [O/H]=[C/H]=-2.5.  The
narrow profile represents thermal broadening, and the broad profile
represents turbulent broadening. 
}
\label{fig:lowZ_systems}
\end{figure}

\begin{figure}
\plotone{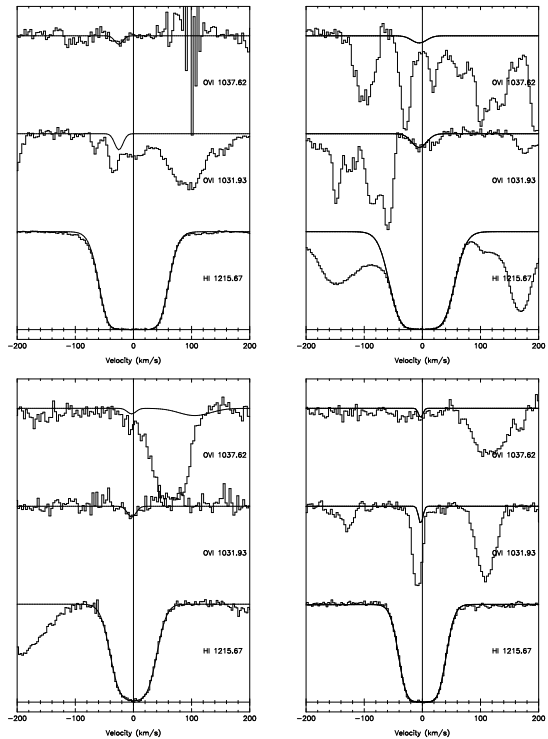}
\caption{ A montage of detected $bd$ systems, shown with best-fit
model curves for the detected component of \ovi.  }
\label{fig:bd_montage}
\end{figure}

\begin{figure}
\plotone{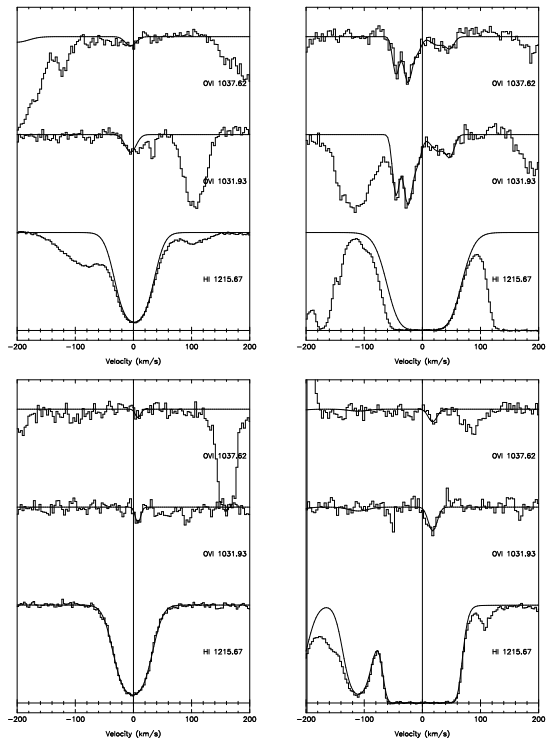}
\caption{ A montage of detected $dd$ systems, shown with best-fit
model curves determined using VPFIT.  }
\label{fig:dd_montage}
\end{figure}

In Figure \ref{fig:lowZ_systems} we show a collection of four low
metallicity systems, to give a sense of the data quality used for our
measurements.  In each case, the \hi profile is shown at bottom, along
with the \civ 1549\AA ~transition and one or both components of \ovi.
We also show with smooth curves the predicted strength of the metal
absorption features for [O/H]=[C/H]=-2.5 and our fiducial ionizing
background spectrum.  Two model curves are shown; the narrower assumes
thermal line broadening and the wider assumes pure turbulence.  In
each of these systems \civ should be marginally detectable, and \ovi
would have been clearly seen.  For comparison, in Figures
\ref{fig:bd_montage} and \ref{fig:dd_montage} we show examples from
the $bd$ and $dd$ samples (respectively), along with their best-fit
absorption line models.

The qualitative results of this section are not strongly sensitive to
the choice of ionizing background spectrum, though there are some
slight differences between the models which are discussed further in
Section \ref{sec:pixels}.  Harder UV backgrounds lead to a slightly
lower average [O/H] for the detected systems (by 0.1-0.2 dex), but
there is still no strong trend of metallicity with \hi column density.
The metallicity-density relation {\em is} affected by the strength of
the soft X-Ray background, in the sense that fewer X-Rays (i.e. a
softer overall background spectrum) would introduce a correlation of
decreasing metallicity with decreasing density (See Appendix B for a
more detailed discussion of this effect).  The upper limits on [O/H]
for the lowest metallicity systems are not affected by changes in the
X-Ray background, but harder UV spectra can reduce these limits from
$[O/H]\le-3.15$ to $[O/H]\le-3.4$.

In physical terms, the \hi densities probed by the survey extend 
roughly to the boundary between the filaments and voids seen 
in cosmological simulations.  Within the filaments, where most baryons 
are found, the enrichment pattern does not appear to vary with 
overdensity.  If strong overall metallicity gradients do exist in 
the IGM, they must occur as the transition is made into the 
voids which fill most of the cosmic volume.  In the moderate 
density regions where our \ovi measurements are most sensitive, we 
detect significant numbers of metal-poor forest lines 
($[O/H]\le-3.2$), and we hypothesize that analogous low 
metallicity systems are likely to exist at lower
densities.  In the subsequent sections, we use the individual
measurements and upper limits to better quantify the relative
frequency of the enriched versus metal-poor systems in our sample.

\subsection{Survival Analysis}

To translate the measurements of individual lines into a statistical
distribution of metallicity, a proper accounting must be made
both of detections and of nondetections in the total sample.  
This problem is well suited to the methods of survival analysis, a
branch of statistics that deals with ``censored'' datasets, which
contain mixtures of measurements and upper limits.  

The most general single variate survival statistic is the Kaplan-Meier 
product limit, which provides a non-parametric maximum 
likelihood estimate of a distribution directly from observed 
data.  The Kaplan-Meier statistic and other survival methods have been
discussed extensively in the astronomical literature
\citep[e.g.,][]{feigelson_survival,schmitt_survival,wardle_knapp}.  
We follow these
examples, modifying notation slightly to suit our particular
application.  We shall use the term ``measurement'' to describe 
the combined set of detections plus upper limits; ``detection'' and
``upper limit'' will be used when the specific context requires.  
We also use the notation $Z_{\rm true}$ to represent the actual
metallicity of a line, so that for detections $Z_{\rm true}=Z_i$, while
for upper limits its value is unknown beyond that $Z_{\rm true}<Z_i$.  

Consider a sample of lines for which a single variate - the
metallicity $Z$ - has either been detected or an upper limit has been 
determined.  We assume that the total set of measurements 
$Z_i, i=1,2,3...N$ has been sorted in order of increasing metallicity 
such that every $Z_i\le Z_{i+1}$ (where multiple measurements result in 
the same value, censored data points are considered smaller).  We wish
to use this sample to estimate the cumulative probability 
distribution $P(Z\ge Z_i)$: the fraction of \lya forest lines above
any given metallicity threshold.

We begin at the maximum of the distribution, since 
for all $Z^{+}>Z_N$, every measurement in the sample is at lower metallicity,
hence we estimate $P(Z<Z^{+})=1$.  According to standard convention, then
\begin{equation}
P(Z\ge Z^{+})=1-P(Z<Z^{+})=0.
\label{eqn:prob_sum}
\end{equation}

Subsequent values of $P(Z\ge Z_i)$ at are constructed at each $i$ by stepping
downwards from the $N$th measurement using the rules of conditional
probability.  As the first such example, we calculate $P(Z\ge Z_N)$:
\begin{eqnarray}
P(Z\ge Z_N)&=&1-P(Z<Z_N)  \\
 &=&1 - P_{[Z<Z_N|Z<Z^{+}]} \cdot P(Z<Z^{+}) \nonumber \\
 &=&1 - P_{[Z<Z_N|Z<Z^{+}]} \nonumber
\label{eqn:prod_firsttime}
\end{eqnarray}
where $P_{[Z<Z_N|Z<Z^{+}]}$ denotes the conditional probability that
$Z<Z_N$ given that $Z<Z^{+}$.  It follows that $P(Z\ge Z_{N-1})$ becomes 
\begin{eqnarray}
P(Z\ge Z_{N-1})=1 - P_{[Z<Z_{N-1}|Z<Z_{N}]} ~~~~~~~~~~~~\nonumber \\
\cdot P_{[Z<Z_N|Z<Z^{+}]} \cdot P(Z<Z^{+}),
\end{eqnarray}
or for arbitrary $i$,
\begin{equation}
P(Z\ge Z_i) = 1 - \prod_{j=i}^{N} P_{[Z<Z_j|Z<Z_{j+1}]}.
\label{eqn:kapmeier_definition}
\end{equation}

The conditional probabilities are estimated by counting measurements
from the ranked sample as follows.  For illustration, we first assume 
all data points in the sample are detections (i.e. no upper limits). 
In this event the conditional probability at each $i$ would be 
\begin{eqnarray}
P_{[Z<Z_j|Z<Z_{j+1}]}&=&
{{{\rm \# ~detections ~with} ~Z<Z_j}\over
{{\rm \# ~detections ~with} ~Z<Z_{j+1}}} \\
\nonumber \\
&=& {{n_{(Z<Z_{j+1})}-n_{(Z=Z_j)}}\over{n_{(Z< Z_{j+1})}}} ~~~(no ~upper ~limits).\nonumber
\label{eqn:prob_nolimits}
\end{eqnarray}

In a sample containing upper limits, the quantities $n_{(Z<Z_{j+1})}$
and $n_{(Z=Z_j)}$ are not uniquely known, since an upper limit $Z_{k}$
measured at $k\ge j+1$ could have a true metallicity $Z_{k,{\rm
true}}<Z_j<Z_{j+1}$, or $Z_j<Z_{k,{\rm true}}\le Z_{j+1}$, or $Z_j
<Z_{j+1}\le Z_{k,{\rm true}}$ .  In other words, an upper limit
$Z_{k}$ with $k>j$ does not provide useful information about the
relative weights $n_{(Z<Z_{j+1})}$ and $n_{(Z=Z_j)}$ because the
relation between $Z_{k, {\rm true}}$, $Z_j$ and $Z_{j+1}$ is
ambiguous.  The Kaplan-Meier method circumvents this ambiguity by
effectively ignoring upper limits with $k\ge j+1$ in constructing
$P_{[Z<Z_j|Z<Z_{j+1}]}$, but retaining the useful information from
upper limits with $k\le j$.  Accordingly, the conditional probability
$P_{[Z<Z_j|Z<Z_{j+1}]}$ is expressed as
\begin{equation}
{{{\rm \# ~measurements ~which} ~must {\rm ~have} ~Z_{\rm true}<Z_j}\over{j}}.
\label{eqn:prob_limits}
\end{equation}

In practice, this probability can take on one of two values, depending
on whether $Z_j$ is a detection or an upper limit.  If $Z_j$ is an
upper limit, then all of the lines up to {\em and including}
line $j$ must have true metallicities less than $Z_j$, hence
$P_{[Z<Z_j|Z<Z_{j+1}]}=j/j=1$.  If $Z_j$ is a detection, then the
conditional probability resembes equation 7:
\begin{equation}
P_{[Z<Z_j|Z<Z_{j+1}]} = \left\{ \begin{array}{ll}
  1 & \mbox{$Z_j$ = upper limit} \\
  {{j-n_{(Z=Z_j)}}\over{j}} & \mbox {$Z_j$ = detection} \\
  \end{array}
\right.
\label{eqn:cond_prob_definition}
\end{equation}
Thus the Kaplan-Meier product limit estimate of $P(Z\ge Z_i)$,
specified by Equations \ref{eqn:kapmeier_definition} and
\ref{eqn:cond_prob_definition}, is a piecewise step function, which
jumps at the $Z$ values of detections and remains constant through
upper limits.

For the Kaplan-Meier estimator to be valid, two assumptions about the
distribution of upper limits must be satisfied.  First, the upper
limits must be independent of one another.  In our case, this is
clearly true when upper limits are measured for lines that are not
blends of two \ovi or \civ components.  This applies to virtually the
entire sample (In some cases, detections are measured in regions with
blends, but never upper limits).  Second, the probability that a
measurement will be censored should not correlate with the measurement
value itself (i.e. the censoring should be random).  Our data may not
strictly meet this criterion, as very high metallicity points are not
likely to be censored.  Yet the sample was selected based on \hi
column density - not metallicity - and since [O/H] and [C/H] do not
correlate with \hi column density (see Figure
\ref{fig:ovi_metal_scatter}) this selection method should randomize
the censoring.  Furthermore, since the measurements are made in data
with a range of signal-to-noise ratio and blending from the \lya
forest, the censoring will be further randomized by variations in the
data quality.  Thus, while the censoring pattern may not be random at
the outset, several factors conspire to make it approximately random
in the end.  This can be seen in Figure \ref{fig:ovi_metal_scatter},
where upper limits are mixed in with measurements over much of the
sample.

\begin{figure}
\plotone{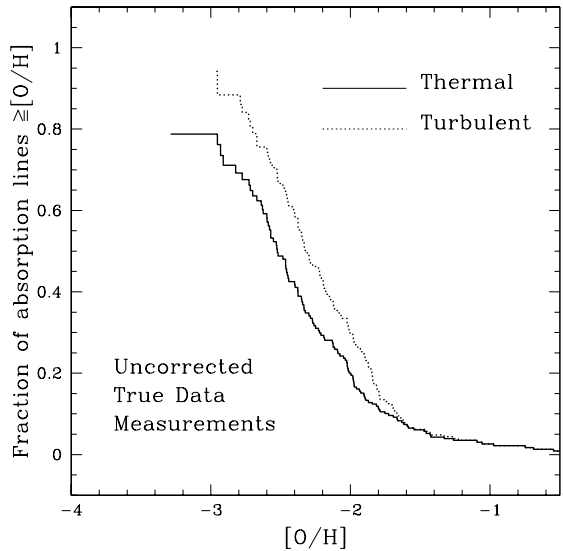}
\caption{ Uncorrected cumulative distribution function of [O/H] for
the \ovi sample.  Both the turbulent and thermal line broadening cases
are shown.  These curves represent upper bounds on the true
distribution, due to contamination from false positive \ovi
identifications. }
\label{fig:kaplan_meier_data}
\end{figure}

We have calculated the Kaplan-Meier distribution for our sample of
metallicities using the publicly available ASURV V1.2 software suite
\citep{asurv_recent,feigelson_survival}.  The raw output for \ovi is
shown in Figure \ref{fig:kaplan_meier_data}.  We ran the calculations
twice, once using our set of upper limit measurements that assume
thermal line broadening (solid line), and once using the limits
measured assuming turbulent broadening (dashed line).  Both curves
have included the entire $dd$ and $bd$ \ovi samples, which will
contain a (possibly large) number of false positive identifications.
As expected, the curve for the turbulent sample is higher than that of
the thermal sample.  This is caused both by the turbulent upper limits
being less sensitive than the thermal upper limits, and by the
introduction of more false positive IDs from statistical fluctuations
in the continuum.  Because of the level of contamination from false
\ovi IDs, the raw Kaplan-Meier distribution represents only an upper
bound on the true metallicity distribution.

Another point of interest is that when the analysis is repeated using
only detected lines (i.e. the upper limits are ignored) the
median metallicity increases by a factor of $\sim 2$, and the
enrichment pattern is similar to what has been inferred 
in earlier studies of intergalactic enrichment
\citep{ellison_civ,cowie_civ_nature,rauch_civ,songaila_civ, 
cowie_civ_2}.  
The median [O/H] in this case is very near $-2.5$, with
scatter of $0.75-1.0$ dex FWHM, and the cumulative probability
distribution is $\ge 90\%$ by [O/H]=-3.0 as though there existed a
metallicity ``floor''.  A proper treatment of upper limits is
therefore a critical component of the analysis, and will have
important implications for the detectability of a universal, zero
point metallicity for the IGM.

\subsection{Corrections for False Positive \ovi Identifications}

Using the tools developed above, we are in position to calculate
correction factors that remove the effect of false positive \ovi
IDs from the data.  We accomplish this using Monte Carlo techniques,
where pairs of \hi and \ovi lines are generated from a predetermined
distribution of [O/H] and added to the data.  We then use
identical measurement methods and survival analyses to test our
ability to recover the input metallicity distribution.  

Our philosophy has been to avoid making corrections directly to the
output of the Kaplan-Meier estimation.  Instead, we note that the 
KM distribution is constructed from a list of 
metallicity detections and upper limits,
where an unknown fraction of the detections are actually false
positives.  For a given detection at risk of being a false positive (e.g.,
a line in the $bd$ sample), the column density that is measured is
unquestionably at least an upper limit on the actual \ovi strength.  
Hence for some unknown fraction of the sample, our
metallicity ``detections'' are actually upper limits on the true 
metallicity.  Our approach has been to use Monte Carlo simulations 
to determine this fraction of false positives.  Then, we create several new
``real'' datasets from the actual data, where different random 
combinations but the same fraction of $bd$ detections are demoted to 
upper limits.  The KM distribution is calculated for each of these 
adjusted samples, and the results are averaged together to obtain our 
final estimate of $P(Z\ge Z_i)$.  

To produce the Monte Carlo data set, we began with the observed $\nh$
and $b_\mhi$ measurements for each quasar sightline, but reassigned a
random redshift to each absorption system.  The redshifts were
distributed according to recent measurements of $dN/dz(z)$ in the \lya
forest in our redshift range \citep{kim_forest}.  For each \hi line we
then generated a corresponding \ovi doublet at the same redshift.  To
determine the $b_\movi$ values, we first assembled the distribution of
$r_b=b_\mhi/b_\movi$ from observed $dd$ systems, where our \ovi
identifications are most secure.  This ratio varies from $r_b=1$ for
turbulent broadening to $r_b=4$ for thermal broadening, with $P(r_b)
\approx 0.5 - 0.07r$ (determined empirically from the $dd$ sample).
No variation in $P(r_b)$ was observed with \hi column density.  Values
of $r_b$ were drawn at random from the distribution and used to
calculate $b_\movi$ for each line in the sample based upon its \hi
linewidth.

\ovi column densities were calculated for each line by inverting
Equation \ref{eqn:metallicity} to find $\novi$ given [O/H] and $\nh$.
The ionization correction term as a function of \hi strength was
calculated with the same HM1.80 ionizing background spectrum used for
the survey data (see Figure \ref{fig:uvbg}).  The actual metallicities
for each line were drawn from an artificial, gaussian distribution
with mean $\left< [O/H] \right> = -2.8$ and $\sigma=0.75$.  This
contrived metallicity distribution was chosen after several iterations
with different backgrounds and metallicity prescriptions.  We found
that the correction factors did not change significantly when other
reasonable backgrounds or input metallicity distributions were used.
Our final background and contrived abundance distribution were chosen
simply to resemble the patterns seen in the actual data.

\begin{figure}
\plottwo{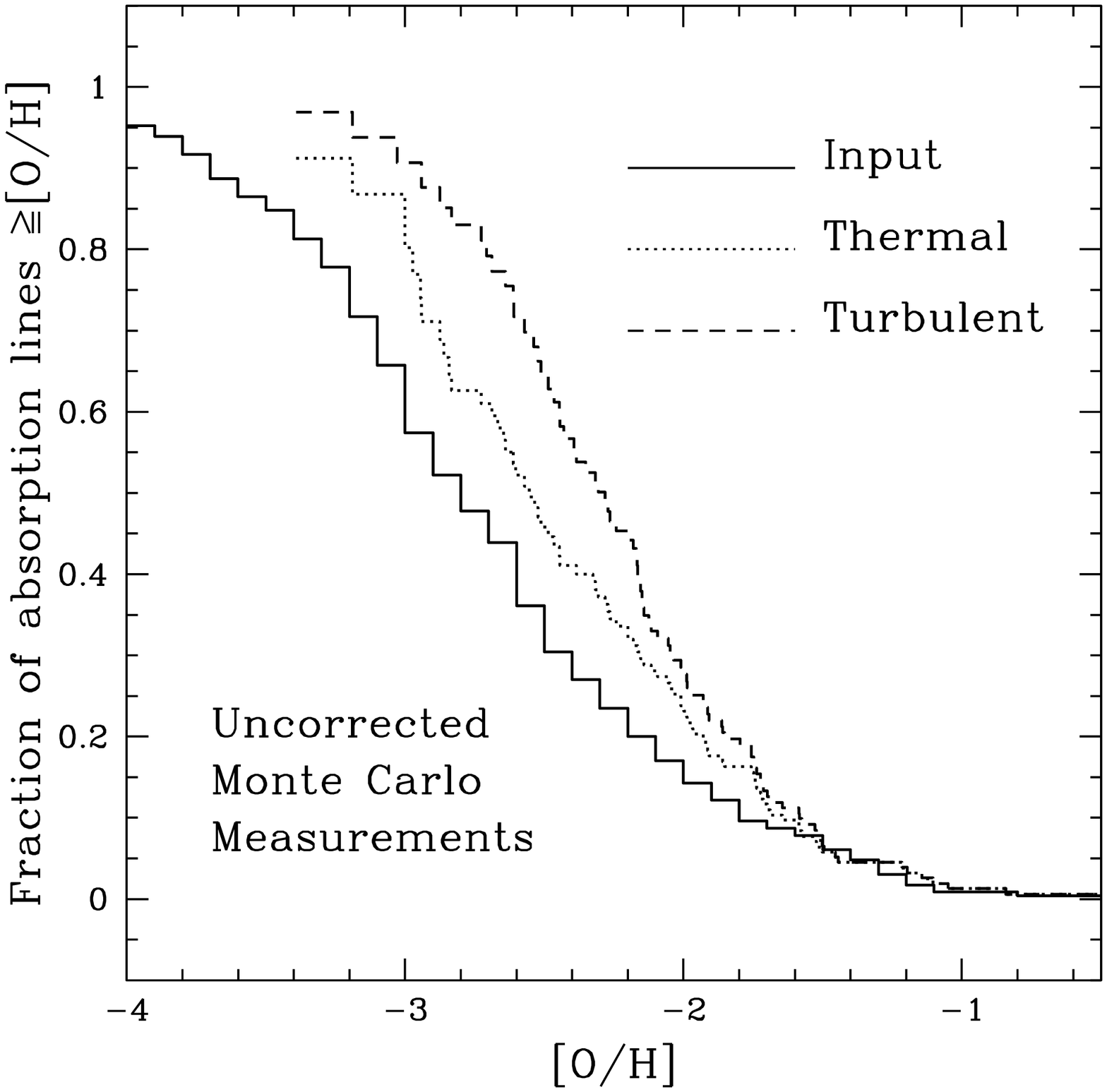}{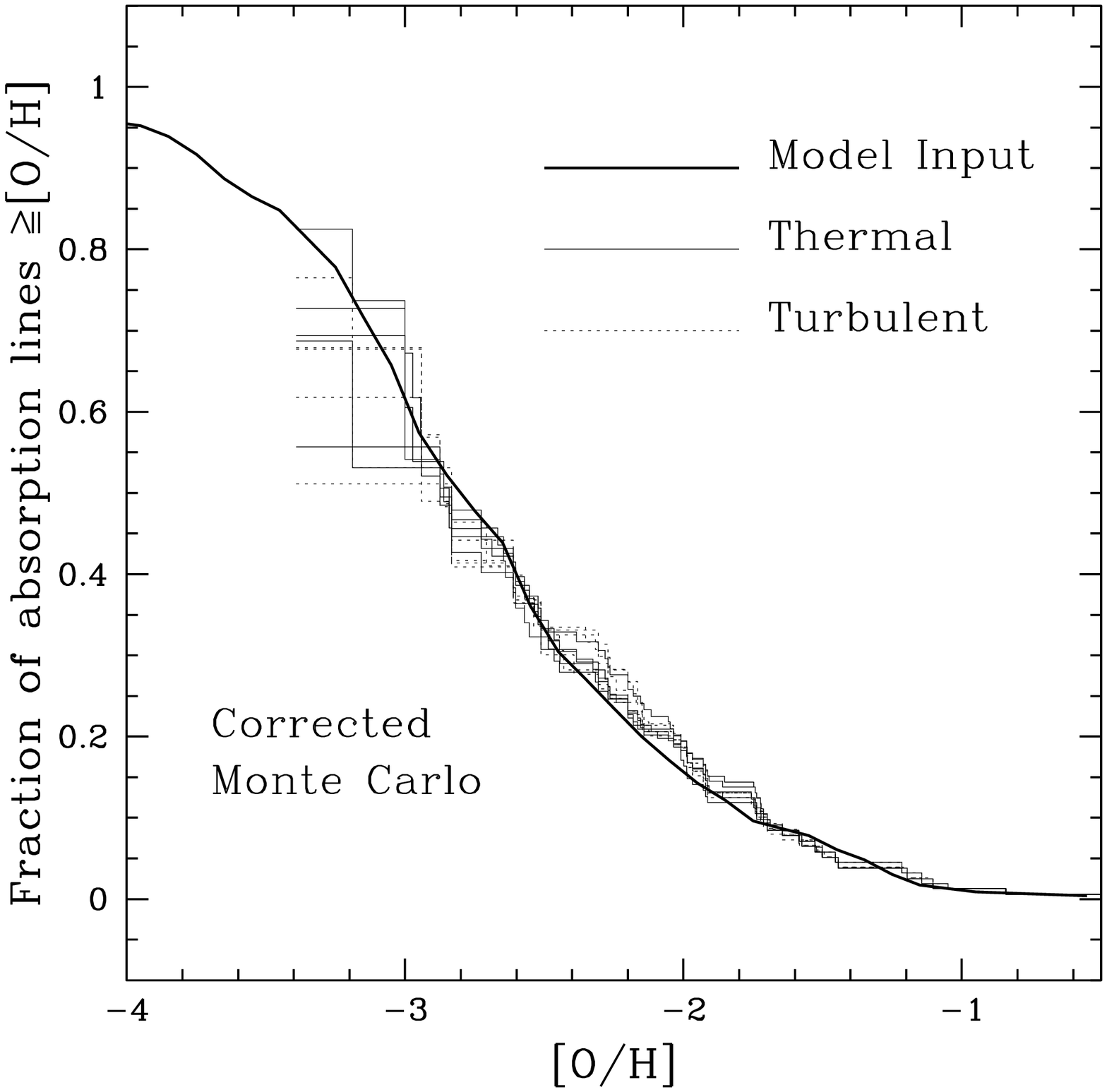}
\caption{ Illustration of the Monte Carlo method to determine
correction factors for false positive \ovi IDs in the data.  The thick
solid line (smoothed in the right panel) indicates the actual
distribution of metallicities put into the simulated spectra.  In the
left panel, we show the raw Kaplan-Meier distributions resulting from
measurements of the simulated lines, in the presence of noise and \lya
forest blending.  The corrected curves in the right panel are created
by randomly re-assigning 57\% of the thermal $bd$ systems as
upper limits (e.g. because of \lya forest contamination).  An $87\%$
correction was required when turbulent line broadening was assumed.
Five realizations of the censoring are shown for each case of the
assumed line broadening, to provide a sense of the random scatter in
the distributions after our correction procedure.  }
\label{fig:montecarlo}
\end{figure}

We added the simulated set of \hi and \ovi lines to the actual
spectra for all 7 quasar sightlines.  This method should best
replicate the actual sources of error and contamination in the real
sample, including continuum fluctuations/errors, varying
signal-to-noise ratio, \lya forest contamination, and instrumental or
other more insidious effects.  To minimize the possibility of
artificial lines being added where real \ovi was already present, none of
the lines were placed at redshifts within 200 km/s of a known line 
with $\nh\ge10^{13.5}$.  The metallicity and \ovi strength were kept
hidden from the user to avoid bias in the detection process.

Finally, we searched the spectra for \ovi absorption at the redshift
of each artificial \hi line in exactly the same manner that was used
for the true data.  We measured \ovi column densities or upper limits
for both the thermal and turbulent broadening cases at each redshift.
These measurements were translated into metallicities, and a
Kaplan-Meier estimate of their distribution function was constructed.
These probability distributions are shown in the left panel of Figure
\ref{fig:montecarlo}.  The thick solid line indicates the true
metallicity distribution that was input to the Monte Carlo line
generator.  The solid and dashed histograms represent the measured
distributions assuming thermal and turbulent line broadening,
respectively.

\begin{figure}[b]
\plotone{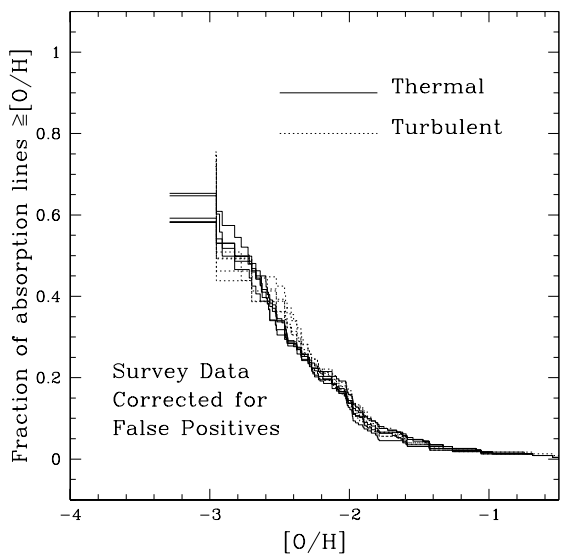}
\caption{ Contamination corrected version of the [O/H] distribution
function for thermal and turbulent broadening, where we have applied
the censoring factors determined from Monte Carlo testing.  Again,
five censoring combinations are shown for each broadening case, as an
indication of the residual scatter from the contamination correction.
}
\label{fig:kaplan_meier_data_corr}
\end{figure}

Clearly the measured distributions systematically overestimate the
input distribution, presumably either because \ovi column densities
have been overestimated due to \lya forest blends, or because false
positive interlopers have contaminated the sample.  We cannot
determine which particular detections are false positives, but the
vast majority will be from the $bd$ sample.  To bring the measured
distributions in Figure \ref{fig:montecarlo} into agreement with the
input distribution, we have demoted varying fractions $f_d$ of the
$bd$ detections to upper limits and recalculated the Kaplan-Meier
distribution.  The specific set of points to be artifically censored
was chosen randomly (for several realizations), and $f_d$ was varied
until the resultant Kaplan-Meier curves produced the closest possible
match to the input distribution in a least-squares sense.  The
exercise was repeated for both the thermal and turbulent
distributions; the final corrected Monte Carlo measurements are shown
in the right panel of Figure \ref{fig:montecarlo}.  Five different
lines apiece are shown for the corrected thermal and turbulent
distributions, corresponding to five realized combinations of the
random censoring.  It was not surprising to find that the false
positive rate was quite high.  For the thermal sample, the optimal
demotion fraction was $f_d=57\%$, whereas for the turbulent sample
$f_d=87\%$.

\subsection{Final Estimates of the Metallicity Distribution}\label{sec:final_Z}

Once the $f_d$ factors had been determined, we applied the same
corrections to the actual survey data.  The results are shown in
Figure \ref{fig:kaplan_meier_data_corr}.  Encouragingly, the corrected
thermal and turbulent distribution functions are in quite close
agreement.  Comparison of Figures \ref{fig:kaplan_meier_data},
\ref{fig:montecarlo}, and \ref{fig:kaplan_meier_data_corr} indicates
that the magnitude and sense of our contamination corrections are
similar for the Monte Carlo and real samples.  This increases our
confidence that the measurement and correction techniques are sound,
at least for similar intrinsic distributions of [O/H].  As a test, we
used the same correction factors to see whether we could recover a
quite different artificial metallicity distribution - a step function
with $P([O/H]<-2.5)=1$ and $P([O/H]>-2.5)=0$.  Again, the Kaplan-Meier
method performed well.  A strong discontinuity was recovered at
$[O/H]\approx-2.45$, though it was broadened by $\sim 0.5$ dex.

\begin{figure}
\plotone{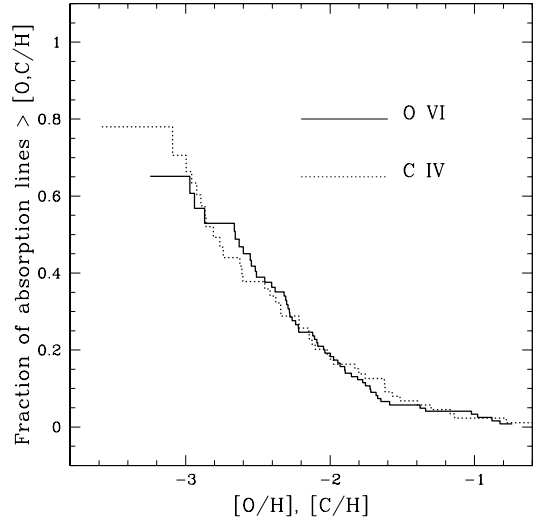}
\caption{Comparison of the Kaplan-Meier distributions for oxygen and
  carbon.  We have assumed our fiducial HM1.80 model for the UV
  background, which was chosen to produce the best possible match
  between these two curves (i.e., we have assumed $[O/C]=0$).  For 
  harder models of the background, \civ is overproduced relative to
  \ovi.  This model produces an excellent statistical match between
  the two distributions.}
\label{fig:km_oc}
\end{figure}

In Figure \ref{fig:km_oc}, we show the Kaplan-Meier distributions of
[O/H] and [C/H] together.  For this figure only, we have restricted
the \ovi systems to have $\nh>10^{14}$, for fair comparison with the
\civ sample.  These curves were produced assuming a HM1.80 UV
background spectrum normalized to $\log J_{912}=-21.5$, yielding a
$95\%$ ionization of cosmic \heii at $z\sim 2.5$.  This choice
produced an excellent statistical match between the [O/H] and [C/H]
curves, and it is used as our fiducial model throughout the paper.  To
arrive at this choice of the background spectrum, we tested a range of
models with different spectral slopes and normalizations to determine
which form produced the best least-squares match between the [O/H] and
[C/H] distributions.  The slopes considered included HM1.50, HM1.55,
HM1.80, and a soft spectrum of HM1.80 quasar light mixed with
radiation from galaxies.  In the QSO+galaxy spectrum, the escape
fraction of Lyman limit photons was set at 10\%.  The normalization of
each spectrum was varied between $-21.6\le\log J_{912}\le-21.2$.

The only spectrum besides the fiducial model that produced
satisfactory results was the HM1.80+galaxy model, normalized to $\log
J_{912}=-21.2$ (implying $93\%$ ionization of \heii).  In this
scenario the [O/H] and [C/H] distributions were both shifted to the
left of those in Figure \ref{fig:km_oc}, by $\sim 0.25$ dex.  The
lower metallicities can be attributed to extra \hi ionizing photons
supplied by the stars.  These boost the hydrogen ionization correction
but do not strongly affect the metal lines, which are governed by
harder UV radiation that can only be supplied by quasars.

In choosing the background spectrum to match [O/H] with [C/H], we have
implicitly assumed that $[C/O]=0$ in the low density IGM, and that the
\citet{grevesse_solar_abund} solar abundances are accurate.  As
discussed in Section \ref{sec:metallicity}, several changes have been
suggested for the solar abundances in recent years.  Substitution of
the \citet{allende_prieto_oc} carbon and oxygen values should not affect
our choice of UV background, since the zero points for these elements
were revised by nearly equal amounts.  However, on the
\citet{holweger_solar_abundances} scale, the solar $(C/O)$ ratio is
increased by 0.16 dex.  Moreover, observations of metal-poor galactic
halo stars indicate that $[C/O]\sim -0.5$ may be a more appropriate
benchmark for low metallicity environments like the \lya forest clouds
\citep{akerman_c_o}.

To gauge the importance of these effects, we used our full suite of UV
background models to compare the distributions of $[O/H]$
vs. $[C/H]+0.5$ (for the metal-poor star scenario), and $[O/H]+0.09$
vs. $[C/H]-0.07$ (to mimic Holweger's solar abundances).  Generally,
we found that harder UV backgrounds resulted in higher [C/O] ratios.
Accordingly, the Holweger abundances favored a HM1.55 background but
ultimately yielded similar intrinsic [O/H] and [C/H] distributions as
our fiducial model.  For all normalizations, the matches between the
Holweger curves were statistically inferior to those of our fiducial
model.  

\begin{figure}
\plotone{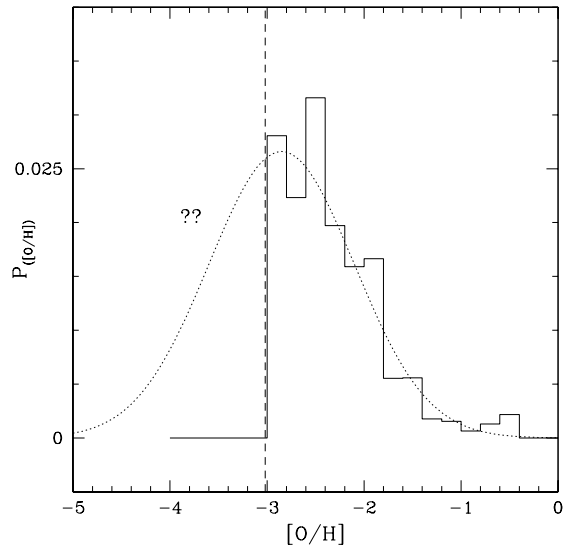}
\caption{The differential distribution function of [O/H], derived from
the Kaplan-Meier estimator.  The dashed vertical line represents the
approximate detection limit of the survey.  The dotted curve is a
unit area gaussian, with mean $\left< [O/H] \right>=-2.85$ and
$\sigma=0.75$, a reasonable analytic estimate of the IGM metallicity to the
limits of our data.  }
\label{fig:diffkm}
\end{figure}

For the $[C/O]=-0.5$ case, a very soft background spectrum was needed
to suppress the measured carbon metallicities relative to oxygen.  In
fact, the HM1.80+galaxy spectrum was the only viable model for any
choice of solar abundances.  Normalizations in the range $-21.6 \le
\log J_{912} \le -21.4$ all produced superb matches between the
distributions of $[O/H]$ and $[C/H]+0.5$.  Thus, {\em to produce an
intergalactic [C/O] ratio resembling that of metal-poor halo stars, a
soft UV background spectrum is required.  In this scenario, the median
carbon abundance is reduced to $[C/H]=-3.1$ and the median oxygen
abundance remains near $[O/H]=-2.7$.}

Figure \ref{fig:kaplan_meier_data_corr} (which includes all \ovi
lines) depicts our best estimate of the metallicity distribution
function of the \lya forest, using our fiducial choice of background
spectrum.  Even at the lowest observed values, the cumulative
distribution does not reach $100\%$.  Roughly $70\%$ of the forest has
been enriched to $[O/H]\ge-3.5$.  Thus {\em we have detected neither a
metallicity ``floor'' in the IGM, nor an abrupt transition toward
chemically pristine gas at low densities}.  An abundance floor could
exist at $[O/H]\le -3.5$, or some of the remaining $30\%$ of lines may
be nearly metal-free.

For our fiducial ionizing background model, we find a median
intergalactic abundance of $[C,O/H]=-2.81$, quite similar to prior
studies suggesting $-2.5\le [O/H]\le -2.0$
\citep{cowie_civ_nature,ellison_civ}, though a factor of $2-4$ lower.
The mean is formally unconstrained, as our lowest measurements are
upper limits.  Some further insight may be gained by examining the
{\em differential} abundance distribution, which we have also
calculated using ASURV using the algorithms presented in
\citet{wardle_knapp}.  The differential \ovi distribution is shown in
Figure \ref{fig:diffkm}, and is quite similar to the \civ distribution
which is not shown.  The vertical dotted line indicates the weakest
detection in the sample (though there are some lower upper limits).
The lines appear to be drawn from a single, fairly uniform
distribution in [O/H] that peaks near the median metallicity.  We
cannot probe the lower half of the distribution as it lies below our
sensitivity threshold, but we know that there must be a substantial
dropoff - since we have already accounted for $\sim 70\%$ of the
lines, the integrated area to the left of the sample cutoff can be at
most $\sim 30\%$.  The observed part of the distribution resembles a
unity normalized gaussian PDF, with mean $\left< [C,O/H] \right> =
-2.85$, and $\sigma=0.75$ dex.  In terms of $Z/Z_\sun$, the
distribution of metallicity would be lognormal.

\begin{figure}
\plotone{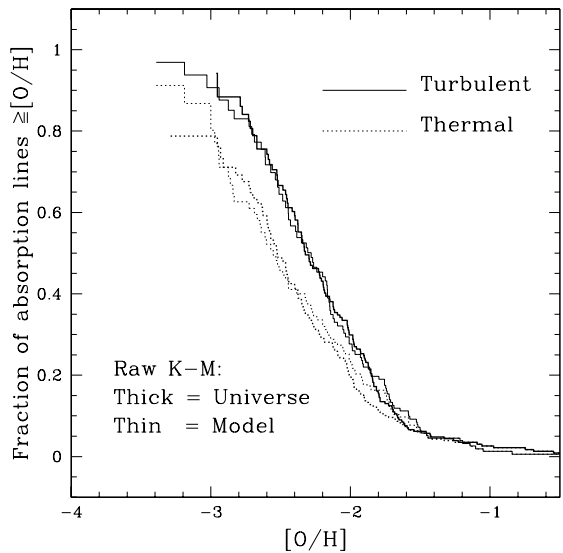}
\caption{Comparison of the raw (uncorrected) [O/H] distribution seen
  in the survey data (thick lines) with the raw distribution produced
  by adding simulated systems to the spectra with our best estimate of
  the differential [O/H] distribution (thin lines).  Our simple
  lognormal metallicity model mimics the real spectra over a broad
  range of [O/H].  }
\label{fig:uncorrected_match}
\end{figure}

To illustrate how our model corresponds to what is seen in the actual
spectra, we show again in Figure \ref{fig:uncorrected_match} the {\em
raw} Kaplan-Meier \ovi distribution, which has not been corrected for
false positive identifications.  Both the thermal and turbulence
broadened measurements are shown, as the thick dotted and solid lines,
respectively.  On the same graph, we have plotted the {\em raw}
Kaplan-Meier distributions of a simulated dataset with the lognormal
metallicity distribution described above (shown as matching thin
curves).  A simple, single population model for the metallicity
distribution (along with a uniform ionizing background) reproduces the
actual survey results extremely well, over the entire measureable
range.

\subsection{Statistical tests for correlation between $\nh$ and [O/H]}

The concepts of survival analysis can also be extended to treat
bivariate data.  This allows us to quantify the trend of [O/H] with
$\nh$, even though the distribution of [O/H] is censored.

Several tests have been devised to rate the significance of
correlations in censored datasets, including the Cox proportional
hazard method, the generalized Kendall's $\tau$ statistic, and
generalized Spearman's $\rho$.  These tests are implemented in ASURV
along with a number of techniques for linear regression of censored
bivariate data.  We have run our measurements through each of the
correlation tests, and performed a Buckley-James regression to
determine the slope of any correlations between [O/H] and $\nh$.  We
used the three different prescriptions for the ionizing background
which produced plausible results in the previous section: our
``fiducial'' HM1.80 quasar spectrum, the HM1.80+galaxy spectrum
normalized to $\log J_{912}=-21.2$ (which produces $[C/O]=0$ but with
lower overall metallicities), and the HM1.80+galaxy spectrum
normalized to $\log J_{912}=-21.5$ (required if $[C/O]=-0.5$).

\begin{figure}
\plotone{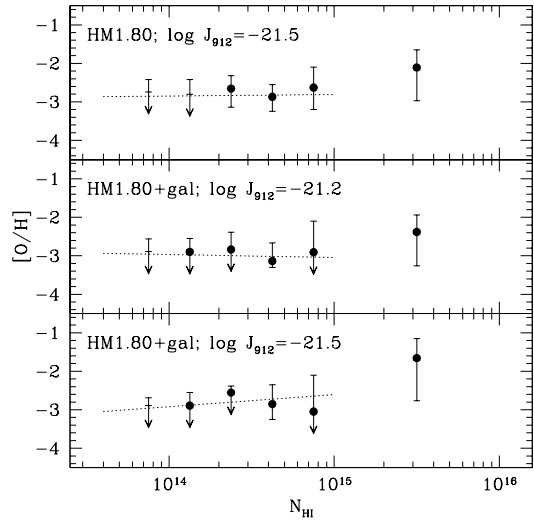}
\caption{Regression analysis to quantify trends of [O/H] as a function
of $\nh$.  Points with vertical bars represent the median [O/H] and
its interquartile range within bins of 0.25 dex in $\nh$, while dotted
lines represent best-fit Buckley-James regression models over the
range $10^{13.6}\le \nh \le 10^{15}$.  For our three favored models of
the UV background, we find no evidence of a strong correlation between
metallicity and overdensity below $\nh\sim 10^{15}$.  At higher
densities we detect a rise in the oxygen abundance, subject to caveats
described in the text.}
\label{fig:regression}
\end{figure}

The results are illustrated in Figure \ref{fig:regression}.  To
produce the data points shown in this plot, we have divided the sample
into bins of $\nh$, and used the Kaplan-Meier statistic to determine
the median [O/H] and its inter-quartile range for each bin.  The
medians are shown as solid dots, and the quartiles are shown as
terminated vertical bars.  For some bins, our measurements were not
sensitive enough to reach the 75th (lowest metallicity) percentile;
these bins are indicated with arrows.  When the distribution
function does not even reach the median for a bin, its lowest reliable
value (typically around the 40th percentile) is indicated with a
horizontal dash.  For each plot, we also show the best-fit relation
between $\nh$ and [O/H], determined using Buckley-James regression
over the range $10^{13.6}\le \nh \le 10^{15}$.  Despite the Figure's
appearance, the line fit is not found though $\chi^2$ minimization of
the binned data.  Its solution more closely resembles a
maximum-likelihood fit to the unbinned points and upper limits.

From the binned data, it is clear that with the $\nh > 10^{15}$
systems included, a significant correlation is present for any choice
of UV background.  This is confirmed by the statistical tests and
regression.  However, we have argued that our metallicities in this
range are less reliable.  Above $\nh= 1-2\times 10^{15}$, a large
fraction of systems no longer conform to our simple, quiescent,
photoionized model of the \lya forest.  In \citet{simcoe2002} we
descibed how these absorbers contain mixtures of hot, probably
collisionally ionized \ovi, together with cooler photoionized \civ and
lower ionization species.  They may be the remnants of galactic winds,
or shocked parcels of unusually metal-rich gas in assembling haloes or
larger structures.  These environments require a more sophisticated
physical model than our simple density and photoionization scalings.
Accordingly, we focus on systems with lower densities for our
regressions.

In the range below $\nh=10^{15}$, there is no obvious signature of a
metallicity gradient in the medians and quartiles, and in fact our
regression fits yielded zero slope for two of our three plausible
background models.  For the fiducial UV background model (top panel of
Figure \ref{fig:regression}), we found a $50-75\%$ probability that no
correlation existed using the Cox method and Kendall's $\tau$.
Spearman's $\rho$ yielded a $88\%$ probability of no correlation.  The
slope of the regression line was $d[O/H]/d\log\nh=0.05\pm 0.13$,
statistically consistent with zero.  The HM1.80+galaxy model
normalized to $\log J_{912}=-21.2$ yielded similar results, though in
this case, the best fit metallicity decreased mildly with increasing
$\nh$ (wih slope $-0.07\pm0.13$).  This represents a physically
unlikely scenario.  There is evidence of a metallicity-density
correlation at the $\sim 2.3\sigma$ level if we use our softest
background, the HM1.80+galaxies with $\log J_{912}=-21.5$ (the
[C/O]=-0.5 condition).  For this UV background the various statistical
tests give a correlation probability between $85-95\%$, with
$d[O/H]/d\log\nh=0.334\pm0.14$.

Our statistical tests support the qualitative results of Section
\ref{sec:metal_scatter}: at low densities we find no strong gradients
in the intergalactic metallicity, though for the softest model of the
UV background that we have tested, there is some evidence of a gradual
trend.  Above $\nh\ge 10^{15}$ ($\oden \sim 11$) there is a clear rise
in the median [O/H], which may mean that a boundary has been crossed
into environments with strong local enrichment.  A corresponding jump
is not seen in [C/H], perhaps indicating that the stronger absorbers
are physically more complex or multiphased.  In the quiescent,
rarefied IGM, the average abundance remains fairly constant with
density to the limits of our survey ($\oden \sim 1.6$).  At the very
least we have not detected a dramatic boundary below which the IGM is
chemically pristine.

\subsection{Comparison with Recent Results from Pixel Statistics}\label{sec:pixels}

A recently published study of \civ by \citet[hereafter
S03]{schaye_civ_pixels} compares observed pixel statistics with
numerically simulated spectra to determine the median intergalactic
carbon abundance and trend of [C/H] with overdensity.  Our results are
consistent with those of S03 where the two studies can be compared,
and the agreement is particularly good regarding the shape of the
metallicity distribution function.  S03 emphasize the presence of a
metallicity-density correlation; in our interpretation of the data
presented above, plausible models of the UV background can lead to
positive, slightly negative, or null correlation results, so we do not
claim to see such compelling evidence of a gradient.

S03 favor an ionizing spectrum similar to our HM1.80+galaxies
prescription, normalized to $\log J_{912}=-21.5$ (at $z\sim 2.5$).
This is the form we used to produce $[C/O]=-0.5$ in the IGM, so
although S03 do not discuss oxygen abundances we expect that their
[O/H] distribution would be higher than the corresponding [C/H].
Using this model we found a median carbon abundance of $[C/H]=-3.1$,
and a median oxygen abundance of $[O/H]=-2.7$.  In S03, the median
carbon abundance increases from $[C/H]=-3.84$ to $-3.19$ over the
range $10^{13.6}\le \nh \le 10^{15}$, with $d[C/H]/d\log\nh\approx
0.43^{+0.07}_{-0.09}$ (note a factor of $\frac{2}{3}$ conversion from
$d [C/H]/d\log\delta$ to $d [C/H]/d\log\nh$, see Equation
\ref{eqn:schaye}).  Our measurements are slightly higher than S03's
over this entire range, although they are comparable near $\nh\sim
10^{15}$.  Our slope for the [O/H] versus $\nh$ fit using this
background is somewhat lower than S03's (0.33 versus 0.43), but still
within the $1\sigma$ errors.  Both we and S03 find lower [C/H] values
than earlier studies \citep[e.g.][]{songaila_civ,ellison_civ}, but
this difference is primarily due to our use of a softer spectrum for
the UV background.

S03 have also tested their calculations using a background spectrum
similar to our fiducial model (their ``Q'' spectrum).  For this case,
they find a higher median $[C/H]\sim -2.8$ to $-2.9$, and little or no
slope in the metallicity-density relation, in good agreement with our
results.  Moreover, both methods derive a lognormal distribution of
metallicities with $\sigma \sim 0.75$ dex.

For the regression analysis, we have taken a slightly different
approach from S03 by dividing our sample at $\nh=10^{15}$, reflecting
our misgivings about the ionization modeling of the stronger systems.
S03 do not split their \civ sample along these lines, though J. Schaye
(private communication) has indicated that their results do not change
substantially if the domain is likewise restricted.  Interestingly, if
we perform a [C/H] regression using the HM1.80+galaxy UV background
with $\log J_{\rm 912}=-21.5$, and all of our \civ systems at all
densities, we obtain a very similar slope to S03:
$d[C/H]/d\log\nh\approx 0.412\pm 0.12$, though the overall
normalization of our [C/H] curve is higher by $0.26$ dex.  The errors
on our [C/H] regression become extremely large if we only consider
systems with $\nh < 10^{15}$, so it is difficult to compare our \civ
results in this regime.  We showed above that our [O/H] regression for
this UV background and density range produced a mild
metallicity-density correlation.  However, a fit for [O/H] including
systems with all \hi column densities produces a much steeper slope
than S03, discrepant at the $3\sigma$ level.

In the end, interpretations of a metallicity gradient are tied to
assumptions about the shape of the UV background and the range of
$\nh$ appropriate to the regression analysis.  We have found three
versions of the UV background that produce plausible relative
distributions of [C/H] and [O/H]; for two of these there is no slope
in the metallicity distribution, and for our third, softest model we
see a mild slope.  Where they can be compared, these results are
fairly consistent with those of S03, who also find little or no slope
for QSO dominated backgrounds, but a slightly larger metallicity
gradient for QSO+galaxy backgrounds.  In no case do we observe a
dramatic decline in the metallicity that would indicate a chemical
enrichment boundary in the IGM.  For all models, the [O/H] slope is
much steeper if we include high column density systems, but in these
cases the slope becomes too severe to be consistent with our low
density result, our [C/H] fit, or the [C/H] fit of S03.  We interpret
this as the onset of a more complicated ionization structure in the
high density systems.

\subsection{Cosmic Mass / Volume Fractions Probed by the Survey}

We now present a simple method for recasting the observed fractions of
absorption lines at different abundances into more physical terms.  We
are particularly interested in the mass and volume fractions of the
universe that are enriched to various levels.  To estimate this
distribution, we first recall that we have fit Voigt profiles to every
\hi line in the \lya forest.  This was originally done to remove
higher order \hi transitions for the \ovi search, but the line lists
also provide detailed information on the column density distribution
of \hi in the exact redshift window of the survey.  This information
may be used to calculate the total mass fraction of baryons probed by
the survey, and by extension the volume fraction.

\begin{figure}
\plotone{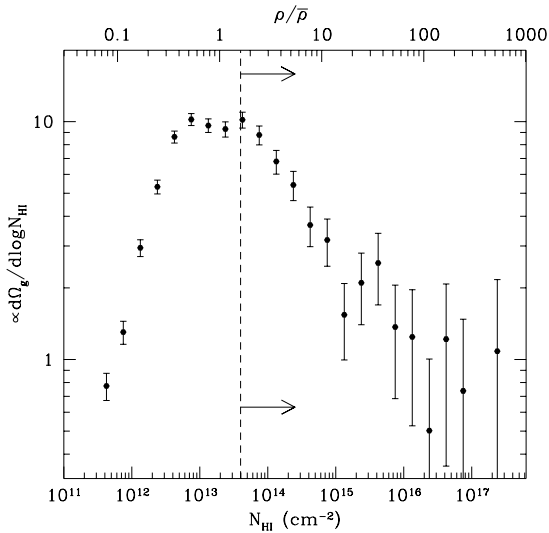}
\caption{The mass fraction of intergalactic gas 
  contained in each decade of \hi column
  density for our line samples.  The vertical dashed line indicates
  the low density cutoff of the \ovi sample, which encompasses $\sim
  50\%$ of $\Omega_g$.}
\label{fig:domega}
\end{figure}

Our mass fraction calculation follows the development of
\citet{schaye_forest}, and we begin with his Equation 16:
\begin{equation}
\Omega_g \propto \int \nh^{1/3} f_{(\nh,z)}d\nh.
\label{eqn:omegag}
\end{equation}
Here, $f_{(\nh,z)}=d^2 {\cal N}/d\nh dX$ represents the number of
\hi absorption lines per unit $\nh$ and per unit absorption pathlength 
\footnote{The absorption pathlength is similar to a redshift interval,
scaled to comoving units.  In a flat universe, a QSO sightline ranging
from $z_{min}$ to $z_{max}$ has an absorption pathlength of $\Delta X
=
\frac{2}{3\Omega_M}\left[\sqrt{\Omega_M(1+z_{max})^3+\Omega_\Lambda}-\sqrt{\Omega_M(1+z_{min})^3+\Omega_\Lambda}\right]$.},
and is calculated directly from the \hi line lists.  We have assumed
(as in Schaye) that the gas is isothermal, and that the fraction of
the total cosmic mass contained in gas is close to
$\Omega_b/\Omega_m$.  Using Equation \ref{eqn:omegag}, we may infer
the fraction of $\Omega_g$ contained in each decade of $\nh$:
\begin{equation}
{{d\Omega_g}\over{d\log \nh}} = {{d\Omega_g}\over{d\nh}}\cdot \nh \propto
\nh^{\frac{4}{3}} \cdot f_{(\nh,z)}.
\label{eqn:domega}
\end{equation}
This quantity is plotted in Figure \ref{fig:domega}, with the points and
errors taken directly from the \hi distribution of the data.  The
vertical normalization of the plot is arbitrary, and the points are
equally spaced in $\log \nh$.  The distribution should be quite
accurate over the range $10^{12.5}<\nh<10^{15}$, where
distinct absorption lines are clearly visible in the data and line 
saturation is not severe.  At higher $\nh$ there is an increased uncertainty in
our column density measurements for some systems.  Since the primary
purpose of the line fits was to facilitate the \ovi search, the
fit quality was not as uniformly high over all \hi transitions for
high density systems that did not have bearing on an \ovi
measurement.  Furthermore, the exact shape of the turnoff below
$\nh\sim 10^{13}$ is the subject of some contention, as it results
from a combination of real, physical effects and observational
incompleteness.  Though we may miss some amount of the mass in this
region, cosmological simulations indicate that there should be a real
and significant drop in the mass density contained in systems below
$\nh\sim 10^{12.5}$.  Whatever the case, the accuracy of the data 
is highest over the range which contains most of the mass.  We shall 
show below that these uncertainties do not change our qualitative results.  

The \hi cutoff of the \ovi sample ($\nh\ge 10^{13.6}$) is shown in
Figure \ref{fig:domega} as a vertical dashed line.  We wish to
determine what fraction of the total gas mass lies to the right of
this line.  A simple inspection suggests that this fraction is close
to one half, which turns out to be remarkably close to the true value.
Since the data are equally spaced in the log, we may simply sum the
points to the right of the line, and normalize by the sum of all
points to determine what fraction of $\Omega_g$ is probed by the
survey.  Depending on whether lines with $\nh>10^{17.5}$ are included
in the calculation (since they suffer from small number statistics),
we find that the survey probes between $51-57\%$ of the total
$\Omega_g$.  Incompleteness at $\nh<10^{12}$ has little effect on the
results, although incompleteness at $10^{12}\le \nh \le10^{12.5}$
could reduce our coverage to $\sim 42\%$ of $\Omega_g$ if we have
undercounted these lines by a factor of 3 (a fairly conservative
estimate).  Accounting for a range of uncertainty in our measurement
of the \hi column density distribution, we estimate that our survey
probes roughly $42-57\%$ of the gas in the universe at $z\sim 2.5$ by
mass, with the most likely value being close to $50\%$.

\begin{figure}
\plotone{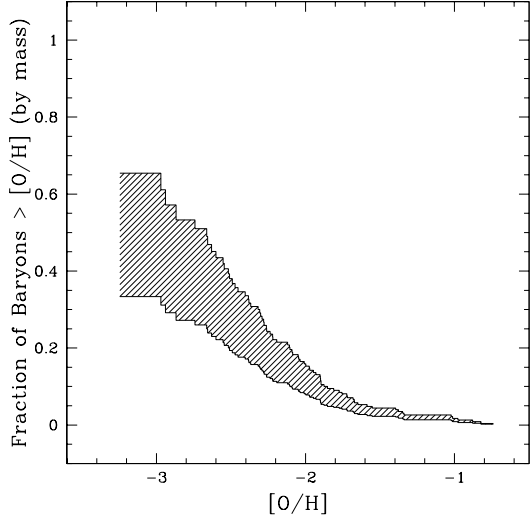}
\caption{ Upper and lower bounds on the enriched mass function of the
IGM, calculated from the \ovi observations.  }
\label{fig:kmmass_ovi}
\end{figure}

In Figure \ref{fig:kmmass_ovi}, we apply this information to determine
the mass fraction of baryons that have been enriched to a given oxygen
abundance, which we call the enriched mass function, or EMF.  The two
curves that bound the shaded region of the plot represent upper and
lower limits on the EMF, derived using two different assumptions about
the trend of [O/H] versus $\nh$ below our survey's \hi threshold.  If
we define $p_{([O/H],\nh)}$ as the differential probability that a
line with $\nh$ will possess an oxygen abundance $[O/H]$, the EMF is
defined as
\begin{eqnarray}
{\rm EMF} & = & f_{{\rm mass}(\ge [O/H])} \\
& = & {\int_{[O/H]}^\infty \int_{-\infty}^{\infty}{
{p_{([O/H],\nh)}{{d\Omega}\over{d\log\nh}}d(\log\nh) 
d \left( \left[ {{O}\over{H}} \right] \right)}} \over
{\int_{-\infty}^{\infty}{{{d\Omega}\over{d\log\nh}}d(\log\nh)}}}.\nonumber
\label{eqn:emf_definition}
\end{eqnarray}
To produce the upper bound, we suppose that the intergalactic
enrichment pattern seen in our data continues to arbitrarily low
densities, filling the entire volume of the universe.  In this case,
$p_{([O/H],\nh)}$ is independent of $\nh$ (See Figure
\ref{fig:ovi_metal_scatter}) and the integrals over $\log\nh$ and
$[O/H]$ may be separated.  The ratio of the density integrals reduces
to unity, and we are left with the expression
\begin{eqnarray}
f_{{\rm mass}{(\ge [O/H])}} & \le & \int_{[O/H]}^\infty{p_{([O/H],\nh)}
d \left( \left[ {{O}\over{H}} \right] \right)} \nonumber \\ 
& = & P_{(\ge [O/H])}.
\label{eqn:emf_upper}
\end{eqnarray}
Thus the upper bound for the EMF is simply the cumulative distribution
of {\em line} enrichment, i.e. the Kaplan-Meier distribution shown in
Figure \ref{fig:kaplan_meier_data_corr}.  We have implicitly assumed that
the lowest density regions of the IGM are not {\em more} heavily
enriched than the highly overdense areas.

To produce the EMF's lower bound, we suppose that all gas in clouds
with $\nh<10^{13.6}$ is chemically pristine, i.e., that we have
already observed all of the heavy elements in the universe.  Now,
Equation \ref{eqn:emf_definition} becomes
\begin{eqnarray}
f_{{\rm mass}{(\ge [O/H])}} & \ge & P_{(\ge [O/H])}\times
{{\int_{13.6}^{\infty}
{{d\Omega}\over{d\log\nh}}d(\log\nh)}\over
{\int_{-\infty}^{\infty}{{{d\Omega}\over{d\log\nh}}d(\log\nh)}}} \nonumber \\
& \approx & 0.5P_{(\ge [O/H])}. 
\label{eqn:emf_lower}
\end{eqnarray}
In the last step, we have used the above estimate that $\sim 50\%$ of
all baryonic mass is in clouds with $\nh\ge 10^{13.6}$.  The upper and
lower bounds on the EMF given by Equations \ref{eqn:emf_upper} and
\ref{eqn:emf_lower} were used to produce the shaded regions in Figure
\ref{fig:kmmass_ovi}.

It is more difficult to estimate the volume filling factor of enriched
gas without reference to simulations, due to the complex nature of the
\lya forest topology, and the lack of a simple analytic form for the
volume fraction $dV/d\nh$.  However, simulations can easily calculate
the volume fraction within a given \hi density contour, or which
contains a certain fraction of $\Omega_b$.  Figure 20b in
\citet{miralda_forest} may be used to estimate the filling factor of
the contour containing half of the baryons at $z=3$.  This number
turns out to be small, only $\sim 5\%$ of the total volume of the
universe.  One can guess why this is the case from Figure
\ref{fig:domega}; we probe to an overdensity of $\rho/\bar{\rho}\sim
1.6$, and most of the mass in the universe is at densities slightly
above the mean (the filaments), while most of the volume is at
densities slightly below the mean (the voids).  We again emphasize,
{\em at the lowest densities where it is possible to probe the
intergalactic metallicity, we only trace cosmic filaments - we have
not reached the voids.}

\subsection{Contribution of Intergalactic Oxygen \& Carbon to Closure Density}

We have estimated the quantity $\Omega_\mciv$ using two different
methods, initially to compare with previous studies and then to test
the consistency of our ionization model.  The first calculation 
takes a direct sum of the observed \civ column densities, as in 
\citet{songaila_omegaz}:
\begin{equation}
\Omega_\mciv = {{1}\over{\rho_c}} m_\mciv {{\sum
    N_{\mciv,i}}\over{\frac{c}{H_0}\sum \Delta X_i}}.
\label{eqn:civ_songaila}
\end{equation}
Here $\rho_c=1.89\times10^{29}$ gm cm$^{-3}$ is the current closure
density, $m_\mciv$ is the mass of the \civ ion, and the upper sum
includes all observed \civ lines.  The lower sum represents the total
absorption pathlength of the sample (defined in the previous section;
note that our formula for $\Delta X$ differs from Songaila's), which
summed over our 2 \civ sightlines comes to $\Delta X = 3.27$.
Applying this to the data we find $\Omega_\mciv=3.50\times
10^{-8}h^{-1}$ (where $H_0=100h$ km s$^{-1}$ Mpc$^{-1}$).  This is in
excellent agreement with other estimates of $\Omega_\mciv$ at $z\sim
2.5$ from the literature \citep{songaila_omegaz,boksenberg}.  In fact,
this value for $\Omega_\mciv$ seems to be observed at all redshifts
out to $z\sim 5$ \citep{songaila_civ,pettini_z5_civ}.  At least for
$z\gtrsim 4$, this may indicate that the \civ systems are enriched
and/or ionized by local sources \citep{boksenberg,pettini_z5_civ}.

\begin{deluxetable}{c c c}
\tablewidth{0pc}
\tablecaption{Metal Density in the Universe}
\tablehead{ {Quantity} & {$\Omega (\times h^{-1})$} & {Method}}

\startdata
$\Omega_\mciv$ & $3.50\times 10^{-8}$ & Eqn. \ref{eqn:civ_songaila} \tablenotemark{1}\\
$\Omega_\mciv$ & $1.31\times 10^{-8}$ & Eqn. \ref{eqn:civ_songaila} \tablenotemark{1}\\
$\Omega_\mciv$ & $1.78\times 10^{-8}$ & Eqn. \ref{eqn:omega_civ_forest} \\
$\Omega_\movi$ & $7.89\times 10^{-8}$ & Eqn. \ref{eqn:omega_civ_forest} \\
$\Omega_{\rm Carbon}$ & $5.10\times 10^{-7}$ & Eqn. \ref{eqn:omega_civ_forest}\tablenotemark{2} \\
$\Omega_{\rm Oxygen}$ & $1.38\times 10^{-6}$ & Eqn. \ref{eqn:omega_civ_forest}\tablenotemark{2} \\
\enddata
\tablenotetext{1}{The first row was calculated using the entire \civ
  sample; the second omits the three strongest lines (out of 83) which
  contain over half of the mass and may not be representative of the 
  intergalactic distribution (see text).}
\tablenotetext{2}{Calculated using Eqn. \ref{eqn:omega_civ_forest}, but with the ionization
  correction factors $f_\mciv$ or $f_\movi$ omitted.}

\end{deluxetable}

\label{tab_omegas}

With a benchmark in hand, we now recalculate $\Omega_\mciv$, this time
starting from the observed distribution of \hi in the forest, and
applying our metallicity and ionization models to predict the total
amount of \civ.  A comparison of this result with the straightforward
sum from Equation \ref{eqn:civ_songaila} tests the consistency of the
enrichment model.  We begin with the \hi column density distribution
function $f_{(\nh,z)}$, calculated from our \hi line fits as described
in the previous section.  We may perform a weighted integral of
$f_{(\nh,z)}$ to obtain an analogue of the last expression in Equation
\ref{eqn:civ_songaila}, with \hi substituted for \civ:
\begin{equation}
{{\sum \nh}\over{\frac{c}{H_0}\sum \Delta X}} \approx
\left( {{c}\over{H_0}}\right)^{-1}\int \nh f_{(\nh,z)}d\nh.
\label{eqn:approx_sum}
\end{equation}
Folding in metallicity and
ionization corrections, we derive the \civ mass density, again through
analogy with Equation \ref{eqn:civ_songaila}:
\begin{eqnarray}
\Omega_\mciv = {{1}\over{\rho_c}} m_\mciv \times 
{
 \left({{C}\over{H}}\right)_\sun \cdot
 \left<{10^{[C/H]}}\right> \cdot
 \left({{c}\over{H_0}}\right)^{-1}  } \nonumber \\
{ \times {\int \nh f_{(X,\nh)}{{f_\mciv}\over{f_\mhi}} d\nh~~~} 
} 
\label{eqn:omega_civ_forest}
\end{eqnarray}
The ionization correction factors, taken from our CLOUDY calculations
(and shown in Figure \ref{fig:ion_fracs}), are density dependent and
therefore included in the integral.  We assume the metallicity
distribution is independent of $\nh$, after Figure
\ref{fig:regression}.  To calculate the mean metallicity, we assume
the [C/H] and [O/H] distributions are identical, and use the
approximation for the [O/H] distribution described Section 3.6 - a
gaussian with mean $\left< \left[ \frac{O}{H}\right]\right>=-2.85$ and
$\sigma=0.75$.  The distribution of $10^{[C/H]}$ is therefore
lognormal, with mean
\begin{eqnarray}
\left<{10^{[C/H]}}\right> & = & \exp{\left[ {\ln 10\times \left<
 \left[\frac{C}{H}\right]\right> + \frac{1}{2}(\ln 10 \times \sigma)^2}
 \right]} \nonumber \\
& = & 10^{-2.20}.
\label{eqn:lognorm_mean}
\end{eqnarray}
Using these values in Equation \ref{eqn:omega_civ_forest}, we find
$\Omega_\mciv=1.78\times 10^{-8}h^{-1}$, similar to the value found
from Equation \ref{eqn:civ_songaila}, but about a factor of 2 lower.
This discrepancy is less significant than it may first appear, since
the sum in Equation \ref{eqn:civ_songaila} is always dominated by the
few strongest systems in the sample.  In our case, three strong
outliers from a sample of 83 lines contain over half of the total \civ
column density.  These are clearly seen in Figure \ref{fig:civhi};
they are found within very strong \civ systems that probably do not
represent tenuous intergalactic matter.  A re-evaluation of Equation
\ref{eqn:civ_songaila} without these three lines included yields
$\Omega_\mciv=1.31\times 10^{-8}$, bringing the estimates from both
methods into close agreement.  This also suggests that a major portion
of the \civ mass in the universe may be contained in a small number of
systems that are highly enriched or harbor unusual ionization
conditions.

When estimating $\Omega_\movi$, it is considerably more difficult to
perform the calculation in Equation \ref{eqn:civ_songaila} because of
\lya forest blending and false positive identifications.  We have
accounted for these effects in our determination of the metallicity
distribution, so it is much more strightforward to apply Equation
\ref{eqn:omega_civ_forest}, yielding $\Omega_\movi=7.89\times
10^{-8}$.  By omitting the \ovi or \civ ionization corrections from
Equation \ref{eqn:omega_civ_forest} (but retaining the \hi correction)
we may estimate the total contribution of carbon or oxygen in all
ionization states to the closure density.  The resulting values are
listed in Table \ref{tab_omegas} along with the values for the \civ
and \ovi ionization states.  \citet{wasserburg_vms} have constructed
an observationally motivated enrichment model for the galactic halo,
requiring several generations of supernovae from very massive stars.
Their prediction for $\Omega_\mciv$ from this model is an order of
magnitude lower than the estimate presented here, and their
$\Omega_\movi$ is 1.65 times larger than our measurement.  We caution
the reader that estimates of [C/O] from our measurements will not be
meaningful, as we have constrained our models of the ionizing
background using assumptions about the [C/O] ratio.

Equation \ref{eqn:omega_civ_forest} is of further interest because it
can be used to determine what fraction of the total \civ (or \ovi
mass) in the universe has been detected in the survey.  This is
achieved by changing the limits of the \hi integration to match our
column density cutoffs - $\nh\ge10^{13.6}$ for \ovi and
$\nh\ge10^{14.0}$ for \civ.  According to this method, our limits
probe 98.8\% of the \civ mass and 90\% of the \ovi mass.

\section{Discussion}\label{sec_discussion}

We have shown that cosmological filaments are enriched with oxygen 
and carbon to near the mean gas density in the universe, and that the
median $[C,O/H]=-2.82$ is similar to previous estimates, though about
a factor of 2 lower.  We have also shown several systems with very low
abundances, implying that $\sim 30\%$ of \lya forest lines 
have metallicities of $[C,O/H]\lesssim -3.5$.  In this section, we
examine our results in the context of cosmic chemical evolution models,
to assess their relevance for distinguishing between
different enrichment mechanisms.

\begin{figure}
\plotone{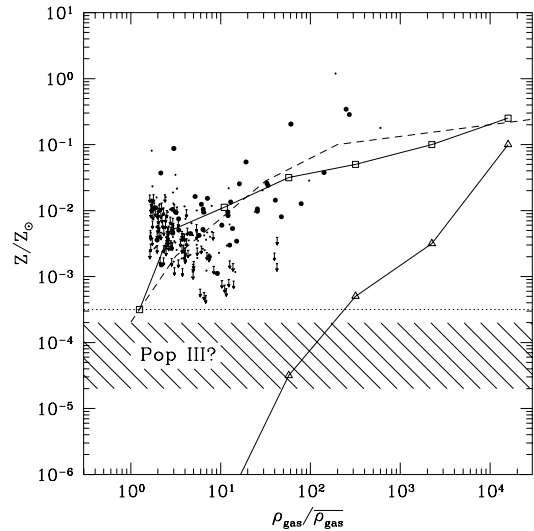}
\caption{ Oxygen abundances for individual \lya forest lines, plotted
against their gas overdensity derived from $\nh$.  Data from the $dd$
and $bd$ sample are shown as heavy and light solid points, though
error bars have been suppressed for clarity.  $3\sigma$ upper limits
are shown with arrows.  The continuous curves depict predictions for
several numerical calculations of the intergalactic enrichment, based
on phenomenological models of star formation and metal ejection.  The
open squares and triangles represent $z=2.3$ models from
\citet{springel_outflows} with and without winds, respectively.  The
dashed line is taken from the $z=3$ wind model of
\citet{aguirre_outflows}.  The shaded region encompasses a possible
range of enrichment from $H_2$ cooled Population III stars, while the
horizontal dotted line represents an absolute upper limit on the
metallicity of these objects \citep[also see
text]{bromm_vms_cutoff_metallicity}.  }
\label{fig:enrichment_models}
\end{figure}

In Figure \ref{fig:enrichment_models}, we show a variant of the
metallicity-density relation from the \ovi sample, now recast in terms
typically used in numerical simulations.  Points from the $dd$ sample
are shown with large symbols; the $bd$ systems are shown with small
dots, $\sim 60\%$ of which are false positives.  The error bars on
detected points have been omitted for clarity.  Along with the data
points, we have indicated the expected metallicity-density relations
for different chemical enrichment scenarios.  We consider two classes
of models: one where metals are expelled at very early times by the
first generation of stars, and one where the ejection epoch occurs
much later and is associated with bursts of ``normal'' star formation
and galactic winds.

Recent studies of star formation in zero metallicity environments show
that the first stars in the universe probably formed via $H_2$ cooling
\citep{abel_first_star,bromm_first_stars,ostriker_gnedin}.  These
``Population III'' stars were uniformly massive ($M\sim
150-250M_\sun$) and short lived, culminating in pair-instability
supernovae that disrupted the entire mass of the star, leaving no
remnant \citep{heger_woosley_vms_yields}.  Yield calculations by these
authors indicate that fully half of the total stellar mass can be
converted into metals and expelled.  All of this takes place prior to
$z\sim30$, at which point the stars are in comparatively small haloes
that are close together.  Thus the Population III objects provide a
potentially efficient mechanism to enrich a large fraction of the
cosmic volume.  However, this mode of star formation is also
self-regulating, which leads to a natural limit to the quantity of
metals that can be produced.  When the first Population III objects
form, they shine brightly in the near ultraviolet and emit enough
photons in the Lyman Werner bands to photodissociate all of the nearby
$H_2$, turning off the cooling source required to produce more stars.
Thus the first major epoch of star formation and metal enrichment is
thought to end at $z\sim 25-30$ \citep{mackey_metals}.

The resulting contribution to the intergalactic metal abundance
depends on fairly uncertain models of the star formation rate prior to
$z\sim 30$ and the effeciency with which the supernova ejecta can
escape protogalactic haloes and mix with distributed gas.  In the
Figure we have shaded a region that reflects current estimates of the
Population III metal production, for mixing efficiencies between $5\%$
and $50\%$ \citep{mackey_metals}.  The horizontal dashed line
represents a more strict upper limit to the Population III enrichment
level.  \citet{bromm_vms_cutoff_metallicity} have shown that massive,
$H_2$ cooled stars will not form from gas with metallicity above this
value, as atomic line cooling becomes too efficient to maintain large
scale stability, and a collapsing cloud fragments into smaller mass
units.

It is clear that the metals we have detected at
$\rho/\bar{\rho}\gtrsim 1$ are in excess of the Population III
production levels, in some cases by several orders of magnitude.  Some
of our most sensitive upper limits are within a factor of 2 of the
model predictions (recall that these are $3\sigma$ limits), indicating
that parts of the IGM could still possess Population III enriched gas.
It is also intriguing that an extrapolation of the Kaplan-Meier
cumulative metallicity distribution (Figure
\ref{fig:kaplan_meier_data_corr}) might be consistent with a universal
floor just below $[C,O/H]=-4.0$, about where one might expect to pick
up Population III metals.  This speculation is neither ruled out, nor
required by the data.

We now turn to chemical enrichment models based on more recent star
formation and galactic winds.  This mode of metal production is more
familiar, as starburst galaxies and superwinds have been observed
extensively at low and high ($z\lesssim 5$) redshifts
\citep{starburst_review,lbg_winds,spinrad_highz_wind}.  The stellar
physics and populations basically resemble those in the local
neighborhood, except that at $z\gtrsim 2.5$ the star formation rate in
the typical galaxy was much higher, and the IMF may have been tilted
towards high mass \citep{cb58_pettini}.  The physics of superwinds is
extremely complex, and a detailed understanding of their impact on the
IGM requires numerical simulations that draw heavily on
observationally motivated recipes for star formation, metal yields,
and wind energetics.  Once these ingredients are in place, the
simulations provide an excellent view of how ejecta interact and mix
with the surrounding IGM.

We have shown the predicted metallicity-density relations from 3
different simulations with continuous curves in Figure
\ref{fig:enrichment_models}.  The solid line connecting open triangles
represents a simulation from \citet{springel_outflows} that includes
local metal enrichment but no galactic winds.  The solid line with
open squares represents a separate run by the same authors, this time
with a fiducial wind model tuned to match observations at low
redshift.  Both curves are for $z=2.3$.  The dashed curve shows the
``W128'' simulation of \citet{aguirre_outflows} at $z=3$, which was
calculated by applying a wind model to already completed numerical
simulations.  Again, the star formation and feedback parameters were
adjusted to match local observations.

We confirm these authors' finding that models with galactic winds are
able to enrich the universe to the levels required by observations.
Moreover, the model without winds grossly underestimates the cosmic
metallicity, implying that {\em essentially all of the heavy elements
yet observed in the IGM were expelled from galaxies at relatively
recent epochs}, loosely $3\lesssim z \lesssim 10$.  This
interpretation is consistent with measurements of turbulence in \civ
systems at similar redshift, which indicate that metal-rich gas is
often stirred on $10-100$ Myr timescales \citep{rauch_civ_lens}.

Both wind models predict a mild correlation of increasing $Z/Z_\sun$
with density, and there is a weak indication that this trend follows
the data.  However, we again caution that our metallicity measurements
are less accurate at $\rho/\bar{\rho}>100$ (see Sections 3.1,3.2), and
the correlation is very weak at lower densities.  No such trend is
seen in the \civ data.  Many of the absorption lines in the range
$\rho/\bar{\rho}=100-300$ may in fact represent snapshots of the
evolving winds themselves, in which case our photoionization models
will probably overestimate [O/H].  These absorbers are complex and
multiphased, but their abundances have been crudely estimated at
$[O/H]\gtrsim -1.5$ \citep{simcoe2002}, roughly consistent with the
models.

Many of our points and upper limits fall well {\em below} the wind
model predictions.  This is most clearly seen at $\rho/\bar{\rho}\sim
8-50$ and may extend to lower densities; the lack of measurements at
$\rho/\bar{\rho}\sim 1-2$ and $Z/Z_\sun\le 10^{-3}$ is a selection
effect, caused by the signal-to-noise dependent detection thresholds
of the data.  Unfortunately, our data are not sensitive enough to
probe $\rho/\bar{\rho}\le 1$, where both wind models predict sharp
abundance falloffs.  With the large parameter space spanned by the
wind and no-wind models, it seems likely that any plausible abundance
pattern could be reproduced given a proper distribution of wind
ejection velocities and metal content.  The model curves represent a
volume average of $Z/Z_\sun$ at different densities; it will be
interesting to test the dispersion in simulated metallicities with
density against the data, though this is beyond the scope of this
paper.

It appears Population III derived metals do not dominate the
intergalactic distribution until one enters the underdense IGM.  Even
then, their spatial variation could be strongly affected if Population
III star formation is strongly biased.  It will be difficult to probe
these metals using direct \ovi measurements, as the fluctuations are
at the $\sim 1\%$ level from the continuum.  Larger telescopes with
very stable instrumentation will be required, and even then the
effects of the \lya forest may render such measurements impossible.  A
more promising approach for the far future will be to search for
intergalactic \ovii and \oviii absorption, as these are the dominant
ionization states in the voids.  However, the technical challenges
involved in observing weak \ovii and \oviii at high redshift will
still be major obstacles for many years to come.

\subsection{The Ultimate Closed Box?}

We now consider a simple, global model of the chemical evolution of the
universe.  This model is analogous to chemical models of the galactic
ISM, but rather than focusing on the transport of mass and chemicals between 
stars and the ISM, we study the transport of mass and chemicals
between galaxies and the IGM, to determine the average chemical yield
of galaxies in the early universe.  As in galactic chemical evolution 
models, we assume
that metals are recycled instantaneously, so for each $dM_{\rm gal}$
of mass flowing into galaxies, a corresponding mass $ydM_{\rm gal}$ of
heavy elements is immediately returned to the IGM.  The 
quantity $y$ is therefore a heavy element yield for galaxy formation.  The
instantaneous approximation should work marginally well for the IGM, 
since the timescale for massive stellar evolution ($\tau\lesssim 10^7$
years for a $M\gtrsim 15M_\sun$ star) and galactic 
winds \citep[$\tau \sim 10^7-10^8$ years;][]{alice_lbg_nirc} is shorter 
than cosmic timescales ($\tau \sim 10^9$ years) at $z\sim 2.5$.

We define $\phi_{\rm gal}$ as the rate of mass deposition into
galaxies from the IGM; this mass may either be converted into 
stars or reside in the ISM of the galaxy.  In each galaxy that is
formed, a fraction $f_{\rm ent}$ of the ISM may be entrained in
outflows and removed from the galaxy before it is substantially
enriched.  The exchange of mass due to galaxy formation is then 
governed by the equations:
\begin{eqnarray}
{{dM_{\rm IGM}}\over{dt}} &=& -(1-f_{\rm ent})\phi_{\rm gal} \label{eqn:closedbox1}\\
{{d}\over{dt}}(ZM_{\rm IGM}) &=& y (1-f_{\rm ent}) \phi_{\rm gal} 
- Z (1-f_{\rm ent}) \phi_{\rm gal}\label{eqn:closedbox2}
\end{eqnarray}
where $M_{\rm IGM}$ is the total mass of gas in the IGM, and as usual
$Z$ is the fraction of $M_{\rm IGM}$ bound up in heavy elements.  By
differentiating Equation \ref{eqn:closedbox2} and dividing by Equation
\ref{eqn:closedbox1}, we obtain
\begin{equation}
M_{\rm IGM}{{dZ}\over{dM_{\rm IGM}}} = -y.
\label{eqn:closedbox_final}
\end{equation}
We have assumed that the total mass $M_{\rm tot}=M_{\rm IGM}+M_{\rm
gal}$ remains constant, i.e., there are no significant sources or
sinks.  We further assume the initial conditions $Z(t_0)=0, M_{\rm
gal}(t_0)=0, M_{\rm gas}(t_0)=M_{\rm tot}$.  This is equivalent to the
``simple closed box'' model of galactic chemical evolution.  Clearly,
the IGM plus galaxies represents the ultimate closed system.

The solution to Equation \ref{eqn:closedbox_final} is
\begin{equation}
Z_{\rm IGM} = y \ln \left({{M_{\rm tot}}\over{M_{\rm IGM}}}\right).
\label{eqn:yield_solution}
\end{equation}
We may use this relation to constrain the metal yield from galaxies
in the era before $z\sim 2.5$.  We have calculated $Z$ from the
metallicity measurements provided above, noting that $Z_{\rm IGM}
=\left< 10^{\rm [O/H]}\right> Z_\sun$, with $\left< 10^{[O/H]}\right> 
\sim 10^{-2.2}$ and
$Z_\sun=0.02$, hence $Z_{\rm IGM}=1.3\times 10^{-4}$.  The
quantities $M_{\rm tot}$ and $M_{\rm IGM}$ may be estimated using
recent observations, first noting that
\begin{equation}
{{M_{\rm tot}}\over{M_{\rm IGM}}} \approx {{\Omega_b}\over{\Omega_{\rm
Ly\alpha}}}
\label{eqn:forest_weight}
\end{equation}
where $\Omega_{\rm Ly\alpha}$ is the contribution to closure density
from diffuse gas in the \lya forest.  The value $\Omega_b h^2=0.0224$ is
now fairly secure from WMAP observations of the CMB
\citep{wmap_params} and Big Bang
Nucleosynthesis observations \citep{omeara_bbn}.  
Our estimation of $\Omega_{\rm
Ly\alpha}$ is somewhat less certain.  A lower bound on this quantity has
been estimated using observations of the average flux decrement in the
\lya forest, either by matching the observations to simulations
\citep{rauch_omegab} or by requiring that the redshift and real-space
extent of \lya forest clouds are of similar magnitude 
\citep{weinberg_omegab}.  Rescaling Weinberg's conservative estimate
to a cosmology with $(\Omega_M,\Omega_\Lambda)=(0.3,0.7)$ gives 
$\Omega_{\rm Ly\alpha}h^2\ge 0.020$ while Rauch finds 
$\Omega_{\rm Ly\alpha}h^2\ge 0.021$.  It seems that there is basic
agreement on a lower limit for the mass of the 
forest: $\Omega_{\rm Ly\alpha}/\Omega_b \gtrsim 0.92$.

We can estimate an upper limit for $\Omega_{\rm Ly\alpha}/\Omega_b$ by
noting that $\Omega_b\approx\Omega_{\rm Ly\alpha}+\Omega_{*}+\Omega_{\rm
DLA} + \Omega_{\rm WH} + \Omega_{\rm hot},$ accounting for most known
sources of baryons in the universe.  The various terms represent the
mass of the \lya forest, stars, damped \lya systems, warm-hot
($T=10^5-10^7$ K) gas, and hot gas in galaxy clusters, respectively.  
To produce a conservative upper limit for $\Omega_{\rm Ly\alpha}$,
we subtract estimates of the mass density in damped systems and stars
from $\Omega_b$.  Of all the constituents of $\Omega_b$, these 
two are the ones most likely to represent local star forming
regions at high redshift.

The value of $\Omega_{\rm DLA}$ has been estimated using extensive
quasar surveys both at high and low resolution.  We use the 
measurements in \citet{storrie_lombardi_omega_dla} for $2<z<3$, 
$\Omega_{\rm DLA}=0.0009\pm0.0003h_{70}^{-1}$.    
The stellar mass density $\Omega_{*}$ is much more difficult to
measure since it is characterized in emission rather than absorption.
\citet{dickinson_omegastar} have used IR photometry to integrate
directly the total $\Omega_{*}$ contained in luminous Lyman break
galaxies from the HDF.  Their photometry corresponds approximately to
rest-frame optical wavelengths, which should be less subject to dust
extinction and/or bursting star formation than rest-frame UV
measurements.  They estimate a conservative lower bound of
$\Omega_{*}\gtrsim 0.0002$ at $z\sim 2.5$, with their best estimate of
the true value being a factor of 5 larger.  Combining the most
conservative estimate of $\Omega_*$ with the measured $\Omega_{\rm
DLA}$, we find $\Omega_{\rm Ly\alpha}\lesssim \Omega_b-\Omega_{\rm
DLA}-\Omega_{*}=0.0446$, leading to the final constraint
\begin{equation}
0.92 \lesssim {{\Omega_{\rm Ly\alpha}}\over{\Omega_b}} \lesssim 0.97,
\label{eqn:forest_limits}
\end{equation}
assuming $h=0.71$.  Entering these values into Equation
\ref{eqn:yield_solution}, the corresponding range for the galactic
yield is
\begin{equation}
0.0015 \lesssim ~y ~\lesssim 0.0041.
\label{eqn:yield_limits}
\end{equation}
In other words, typical galaxies in the early universe recycled
$0.1-0.4\%$ of their input mass back into the IGM as heavy elements.  

It is also interesting to compare the galaxy yield $y$ with the
stellar yield interior to the galaxy, $y_*$.  The stellar yield
specifies the amount of metals released into the galaxy's ISM, $y_*dM_*$
for each $dM_*$ of stars formed.  If we assume that a fraction
$f_{\rm ej}$ of the metals formed in a given starburst are ejected from
the galaxy into the IGM, we may crudely relate the stellar and galactic yields
for a galaxy:
\begin{equation}
y(M_* + M_{\rm ISM}) = y_* M_* f_{\rm ej}.
\label{eqn:stellar_galactic_yields}
\end{equation}
Here, $M_*$ and $M_{\rm ISM}$ represent the amount of mass within the
galaxy locked up in stars and gas, respectively.  These terms may be
rearranged to produce the following expression:
\begin{equation}
f_{\rm ej} = {{y}\over{y_*}}\left( 1 + {{M_{\rm ISM}}\over{M_*}} \right).
\label{eqn:ejection_fraction}
\end{equation}
For bare stars, the ejected metal fraction is simply the ratio of the
galactic and stellar yields; conversely, for galaxies with inefficient
star formation more of the synthesized metals must escape in order to
boost up the galactic yield.  We saw in the previous section that most
of the intergalactic enrichment is caused by relatively recent winds
from star forming galaxies.  Since high redshift galaxies show
evidence for preferential enrichment by Type II supernovae, we
estimate $y_*$ by integrating the supernova yields of
\citet{woosley_weaver_yields} over an initial mass function, from
$0.1M_\sun$ to $100M_\sun$.  We have used various prescriptions for
the IMF, including \citet{salpeter_imf}, \citet{kennicutt_imf}, and
\citet{miller_scalo_imf}, which produce weighted yields ranging from
$y_*=0.001$ (Miller-Scalo) to $y_*=0.011$ (Kennucutt).  The yield in
the solar neighborhood is approximately $y_*\sim 0.02$.  We may set a
conservative lower bound on $f_{\rm ej}$, using $y=0.0015$ (Equation
\ref{eqn:yield_limits}) and $y_*=0.011$, with the result
\begin{equation}
f_{\rm ej} \gtrsim 14\% \left( 1 + {{M_{\rm ISM}}\over{M_*}} \right).
\label{eqn:ejection_numbers}
\end{equation}
The actual value of $f_{\rm ej}$ is probably higher than 14\%, since
at earlier times less of the gas has been converted into stars,
leading to a larger ${{M_{\rm ISM}}\over{M_*}}$.  This could be
mitigated to some extent if the high redshift IMF is extremely
top-heavy, which drives up $y_*$.

Despite our crude models and large uncertainties, the overall
impression is that galaxies at high redshift are quite efficient at
returning the metals they produce to the IGM.  On average, they will
eject over $\frac{1}{10}$ of their nucleosynthetic byproducts.  Such
vigorous recycling suggests that much of the star formation in early
galaxies occurred in bursts, rather than a more quiescent, gradual mode
that would allow them to retain their metals.

Observations of local starbursts indicate that the yields and ejection
fractions we derive are fairly reasonable for star forming dwarf
galaxies.  For example, the nearby dwarf NGC 1569 is driving a
galactic wind that contains $\sim 34,000M_\sun$ of oxygen at $Z\sim
Z_\sun$ \citep{crystal_wind_yield}.  The parent galaxy has a combined
$M_{\rm ISM} + M_* = 1-1.5\times 10^8 M_\sun$.  If we assume a solar
mass fraction of oxygen relative to total heavy element mass
($0.378$), the implied yield for this galaxy is $y=0.0007$ - roughly a
factor of 2 below our lower limit for the high redshift yield.  The
authors also find that the wind carries away nearly all of the metals
produced by the starburst, while the neutral gas disk holds $\sim 5$
times more oxygen from prior periods of quiescent star formation,
implying a metal ejection fraction of $f_{\rm ej}\sim 15-20\%$ over
the lifetime of the galaxy.  We do not wish to overinterpret an
isolated example when discussing the global properties of all galaxies
in the early universe.  We simply note that similar objects at high
redshift could produce and distribute the quantity of heavy elements
seen in the IGM, provided their wind velocities are high enough to
efficiently mix debris over large scales.

\section{Summary and Conclusions}\label{sec_conclusions}

We have presented new observations of \ovi and \civ absorption at
$z\sim 2.5$, which were used to estimate the metallicity distribution
function of the intergalactic medium.  Seven quasar sightlines were
used to measure \ovi for lines with $\nh\ge 10^{13.6}$; two of these
with exceptionally high signal-to-noise ratios were also used to
measure \civ for lines with $\nh\ge 10^{14.0}$.  These limits probe
densities of $\rho/\bar{\rho}\sim 1.6$ relative to the cosmic mean.
For each line in the samples, we estimated [O/H] and [C/H] by applying
density-dependent ionization corrections to the measurements of $\nh,
\novi$, and $\nciv$.  We experimented with several prescriptions for
the ionizing radiation background, eventually settling on a model that
produced the best match between the distributions of [O/H] and [C/H].
This model is dominated by quasar light with an original spectrum of
$F_\nu\propto \nu^{-1.8}$, that has been reprocessed through the IGM,
and normalized to a flux of $\log J=-21.5$ at the \hi Lyman limit.
Motivated by observations of metal-poor galactic halo stars, we also
experimented with the radiation field to produce a relative
intergalactic abundance of [C/O]=-0.5.  A softer UV spectrum including
galactic radiation was required to meet this criterion.

Our sample contains a mixture of detections and upper limits on the
metallicity for each absorption line in the forest.  
The individual \ovi and \civ lines are observed to scatter both above and
below the trend of $[C,O/H]=-2.5$ observed in previous studies of the
intergalactic metallicity, and several upper limits lie nearly an
order of magnitude lower.  Since the sample contains upper limits, we
have used survival statistics to construct a Kaplan-Meier product
limit estimate of the [O/H] and [C/H] distributions within the \lya
forest.  For the \ovi distribution, we carefully corrected for the
effects of false positive identifications that might result from
interloping \hi lines in the \lya forest.  Our basic results may
be summarized as follows: 
\begin{enumerate}
\item{The mixture of heavy elements within cosmological filaments does
  not differ qualitatively in regions of high and low density (for
  roughly $1\lesssim \rho/\bar{\rho} \lesssim 10-15$).  We have not
  observed any significant variation in the [O/H] or [C/H]
  distribution that correlates strongly with $\nh$, though there may
  be a weak correlation when we use our softest model for the ionizing
  background spectrum. }
\item{At $\oden \gtrsim 10-15$, we see evidence of a change in the
  trend of [O/H] relative to [C/H], which we interpret as the onset of
  multi-phased ionization structure in many of the the stronger
  absorbers.}
\item{Roughly $30\%$ of lines in the \lya forest are enriched to
  abundances below $[C,O/H]\lesssim -3.5$.  Thus we have not detected
  evidence for a metallicity floor in the IGM.  If such a floor
  exists, extrapolation of the cumulative abundance distribution
  suggests it would lie in the range $-5\lesssim [C,O/H] \lesssim
  -4$, about a factor of 3 lower than the limits probed by our
  survey.  Nevertheless, some portion of the
  cosmic volume is very metal poor, even in regions with
  $\rho/\bar{\rho}\sim 10$. }
\item{The median abundance of the filaments is $[C,O/H]=-2.82$.  By
  differentiating the cumulative abundance distribution, we obtain the
  probability density for [O/H] and [C/H], though we cannot construct
  the lowest metallicity portion of the distribution with our data.
  For the $\sim 70\%$ of lines which we do measure, the distribution
  may be approximated as a gaussian with mean
  $\left<\left[{{{O}\over{H}}}\right]\right>=-2.85$ and $\sigma=0.75$
  dex.  The distribution of $Z/Z_\sun$ is therefore lognormal, with
  mean $\left<{{Z}\over{Z_\sun}}\right>=10^{-2.2}$.}
\end{enumerate}

It is also instructive to express our results in terms of an enriched
mass function, defined as the cumulative baryonic mass fraction of the
universe that is enriched above a given abundance level.  We place
upper and lower bounds on the EMF by assuming either that the
abundance pattern observed at $\rho/\bar{\rho}\gtrsim 1.6$ extends to
arbitrarily low density, or that we have detected all heavy elements
and that lower density clouds are chemically pristine.  If the true
pattern lies between these two extremes, then $\sim 50-60\%$ of all
baryons in the universe have been enriched to $[C,O/H]\ge -3.5$ by
$z\sim 2.5$.  We have not calculated similar constraints for the
enriched volume fraction, since even at these low column densities we
only probe $\sim 5\%$ of the spatial extent of the universe.
Generally, then, our conclusions apply to the gas found in cosmic
filaments at $z\sim 2.5$, where most of the mass resides.  We have not
yet reached the voids which contain most of the cosmic volume.

We have examined our measurements in light of two established models
for cosmic chemical enrichment.  The first model describes metal
production from the Population III stars postulated to form at $z\ge
30$ from $H_2$ cooling, and the second describes metal-laden
superwinds that are
driven from starburst galaxies at more recent epochs.  A comparison of
these models with our data suggests that Population III stars do not
produce enough metals to enrich the cosmic filaments at observed levels.

Models incorporating galaxy outflows similar to those seen at
low redshift {\em can} reproduce the observed average 
intergalactic metallicity.  The significant scatter observed 
towards low $Z/Z_\sun$ indicates that there may be local variations 
in the frequency or efficiency of these winds.  Yet it seems likely 
that the models will be able to match these trends given proper tuning.
The wind models predict a substantial decrease in the average
metallicity at \hi densities slightly below the threshold of our
survey.  The measurements presented herein do not have sufficient
sensitivity to detect this decline; pixel-statistical methods may improve
on this in the near future, and long term prospects may include
searches for intergalactic \ovii and \oviii, which are the dominant
ionization species at these densities.

To find regions of the universe whose metals are derived from
Population III stars, it will be necessary to survey
the cosmic voids.  However, if Population III star formation is 
significantly biased then it is possible that the underdense IGM
could remain nearly chemically pristine.
The chemical enrichment of cosmological filaments is dominated
by the debris of galactic outflows, and therefore must have taken
place relatively recently.  Taken together with our estimates of the
enriched mass function, this indicates that roughly half of all
baryons have either been processed through a galaxy or mixed with
material that has done so in the first 3 Gyr after the Big Bang.

Finally, we present a simple closed-box model of the chemical
evolution of the universe.  
We use this model along with our metallicity measurements
to estimate the average metal yield of galaxies prior to $z=2.5$.
For a galaxy, the metal yield is determined in part by the star
formation rate and raw stellar yields, but also by the efficiency with
which outflows can expel metals from the galaxy's gravitational
potential.  We find that the typical galaxy ejects $\sim 0.1-0.4\%$ of its
formed mass into the IGM as heavy elements.  This amounts to at least 
$\gtrsim 14\%$ of the total quantity of metals synthesized 
within these galaxies.  These yields are not qualitatively different
from what is seen in some local dwarf starburst galaxies. 

\acknowledgements We gratefully acknowlege the assistance of F. Haardt
in providing us with several different models of the X-Ray/UV
background spectrum in advance of the public release of his CUBA
software package, as well as his graciousness in answering several
questions about the details of the model calculations.  This work
would also not have been possible without several key software
packages which have kindly been made available to the public by their
authors.  These include T. Barlow's MAKEE data reduction software,
R. Carswell's VPFIT, G. Ferland's CLOUDY, and E. Feigelson's ASURV.
We acknowledge helpful readings and/or discussions with M. Pettini,
J. Schaye, and J. Prochaska, and D. Reimers provided advance
information on several interesting targets.  We thank the Keck
Observatory staff for their assistance in carrying out the
observations.  Finally, we wish to extend special thanks to those of
Hawaiian ancestry, for allowing us study the universe from their
sacred mountain.  W.L.W.S. and R.A.S. gratefully acknowledge financial
supprt from NSF grants AST-9900733 and AST-0206067; M.R. is grateful
for support through grant AST-0098492 from the NSF, and grant
AR90213.01A from the Space Telescope Science Institute which is
operated by the Association of Universities for Research in Astronomy
nc., under NASA contract NAS5-26555.

\bibliography{/h0/simcoe/latex/ovi}

\appendix

\section{A Note on Models for the UV/X-Ray Ionizing Background}

All of the models for the metagalactic ionizing background spectrum
were provided to us by F. Haardt, using his CUBA
package.  This software performs calculations similar to those
described in \citet{HM96}, with some new features.  The basic
principle is to integrate the best available redshift-dependent
luminosity functions from the literature to create an average source
spectrum of galaxies and quasars at each redshift.  Then, the photons are
statistically propagated through the a parametrized model of the IGM 
to simulate absorption and reemission at important ionization edges.  
The newer models extend the range of the original \citet{HM96} spectra
by including the X-Ray background, which turns out to be quite
important for \ovi in the low-density IGM.  We have already described
in Section 3.3 some of the uncertainties in the UV portion of the
spectrum (e.g. the range in measurements of the spectral slope $\alpha$) and
their effects on our results.  

The model for the X-ray background is calculated independently from
that of the UV/optical light in the new models.  It is assumed to
originate only from AGN, and its source function is 
calculated by integrating the X-ray luminosity function at each
redshift.  This function is assumed to evolve
independently of the optical/UV LF, and its redshift-dependent 
form is taken from \citet{boyle_xrlf}.  
Seyfert I galaxies contribute flux at all
X-ray wavelengths, since the IGM is exposed to the bare X-ray flux
from the central AGN engine.  The core of Seyfert II galaxies is
covered by an obscuring torus which absorbs soft X-ray photons but
transmits hard X-rays ($E\gtrsim 1$ keV).  The hardness of the X-ray 
background is thus sensitive to the relative numbers of Seyfert I and
II galaxies at each redshift.

The UV/X-ray background models use weighted combinations of Seyfert I
and II X-ray spectra, constructed as described in \citet{madau_xrbg}.
The evolution in number ratio of Seyfert I / Seyfert II galaxies is
estimated by assuming that AGN with B-band luminosity above a given
threshold are always Seyfert I, while those below are Seyfert II.
This is normalized to locally observed Type I/Type II fractions.  The
models match observations of the UV and hard X-ray backgrounds at low
redshift, within errors \citep{haardt_cuba}.  The local soft X-ray
metagalctic background is less well constrained, partly because of
significant galactic foregrounds.  This range turns out to be most
important for our \ovi results.

At $z\sim 2.5$, the UV/optical luminosity function is near its peak
value over the history of the universe.  Since the models calculate
the Type I/Type II AGN fraction based on optical flux, this means that
nearly all the AGN at this redshift ($> 95\%$) are counted as Seyfert
I, and the soft X-ray background is quite high.  The models are still
uncertain, but it seems more likely that the soft X-ray backgrounds
used here would err on the high side rather than the low, since 
essentially no dilution by Type II AGN is included at these redshifts.  
This affects our results in the sense that a higher soft X-ray flux
will suppress \ovi production, as described in Appendix B.  

We have performed the entire analysis for several models of the
ionizing background, retaining the HM1.80 model with $\log
J_{912}=-21.5$ since this produces the best match between the observed
[O/H] and [C/H] distributions.  The other models we tested included
HM1.50 and HM1.55 slopes, with normalizations of $\log J_{912}=-21.1,
-21.4, -21.5,$ and $-21.6$, and an HM1.80 model with \hi ionizing flux
from galaxies included.  Our optimal normalization is slightly lower
than estimates of $J_{912}$ made using the proximity effect at similar
redshifts, though the discrepancy is only at the $\sim 1\sigma$ level.
The field of X-ray population modeling at high redshift is rapidly
evolving with the comprehensive surveys being undertaken by XMM and
Chandra, and much better models of the high redshift X-ray background
should be forthcoming in the near future.

\section{Effects of the X-Ray background on derived Oxygen, Carbon Abundances}

It is seldom appreciated that for densities below $\nh\sim 3\times 10^{14}$, 
the oxygen in the IGM is actually {\em over}ionized relative to \ovi 
- most atoms are in the \ovii and \oviii states (See figure 4).  In 
photoionization equilibrium, the density of \ovi is governed according 
to the equation
\begin{equation}
n_\movi \int_{138 {\rm eV}}^{\infty}{{4\pi J(\nu) \sigma(\nu)}\over{h\nu}}d\nu
= n_e n_\movii\alpha(T),
\label{eqn:ion_balance}
\end{equation}
and likewise for other ionization states.  Since at low densities
$N_{\mov}\ll N_\movii$, much more \ovi is produced by recombination from
\ovii than photoionization of \ov.  Accordingly, the abundance 
of \ovi is determined by the number of \ovii ions available for 
recombination.  At low densities, \ovii and \oviii are found in
similar proportions and their ratio is sensitive to the background
flux at the \ovii$\rightarrow$\oviii ionization edge.  This ionization
energy is in the soft X-Ray band (739eV, or 17\AA, indicated 
in Figure \ref{fig:uvbg}), which leads to the counterintuitive result that the
intergalactic \ovi abundance is critically sensitive to the intensity
of the soft X-Ray background at low densities.  A decrease in the 
soft X-Ray background favors a higher \ovii abundance, and hence 
actually {\em enhances} \ovi.  

At densities above $\nh\sim 3\times 10^{14}$, most oxygen is in either
the \ovi or \ovii state.  The ionic balance is then governed by the
intensity of the UV background at the \ov$\rightarrow$\ovi and
\ovi$\rightarrow$\ovii ionization edges (109.5\AA ~and 89.5 \AA,
indicated in Figure \ref{fig:uvbg}).  In this regime, a decrease in
the background results in fewer photons with sufficient energy to
create \ovi, thus lowering the predicted \ovi column densities.  Thus
a substantial decrease in the X-ray background would cause our
metallicities for the lowest density lines to move to lower values,
but leave the higher density lines basically unchanged - introducing a
dropoff in metallicity below $\nh\sim 10^{14}$.  Our best current
models do not reveal such a trend, but the reader should bear in mind
that a significant revision of the X-ray background at $z\sim 2.5$
could affect this result.


\end{document}